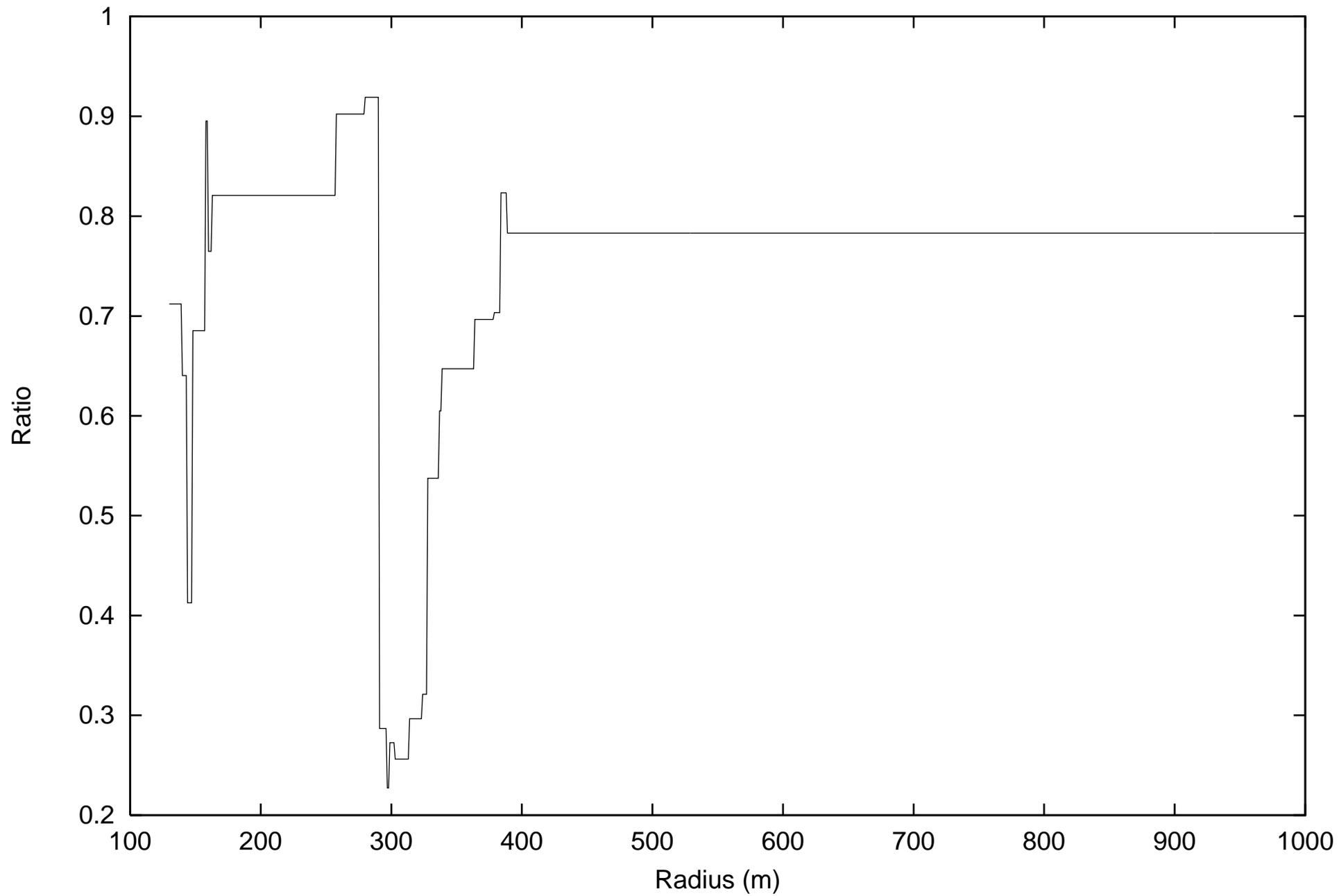

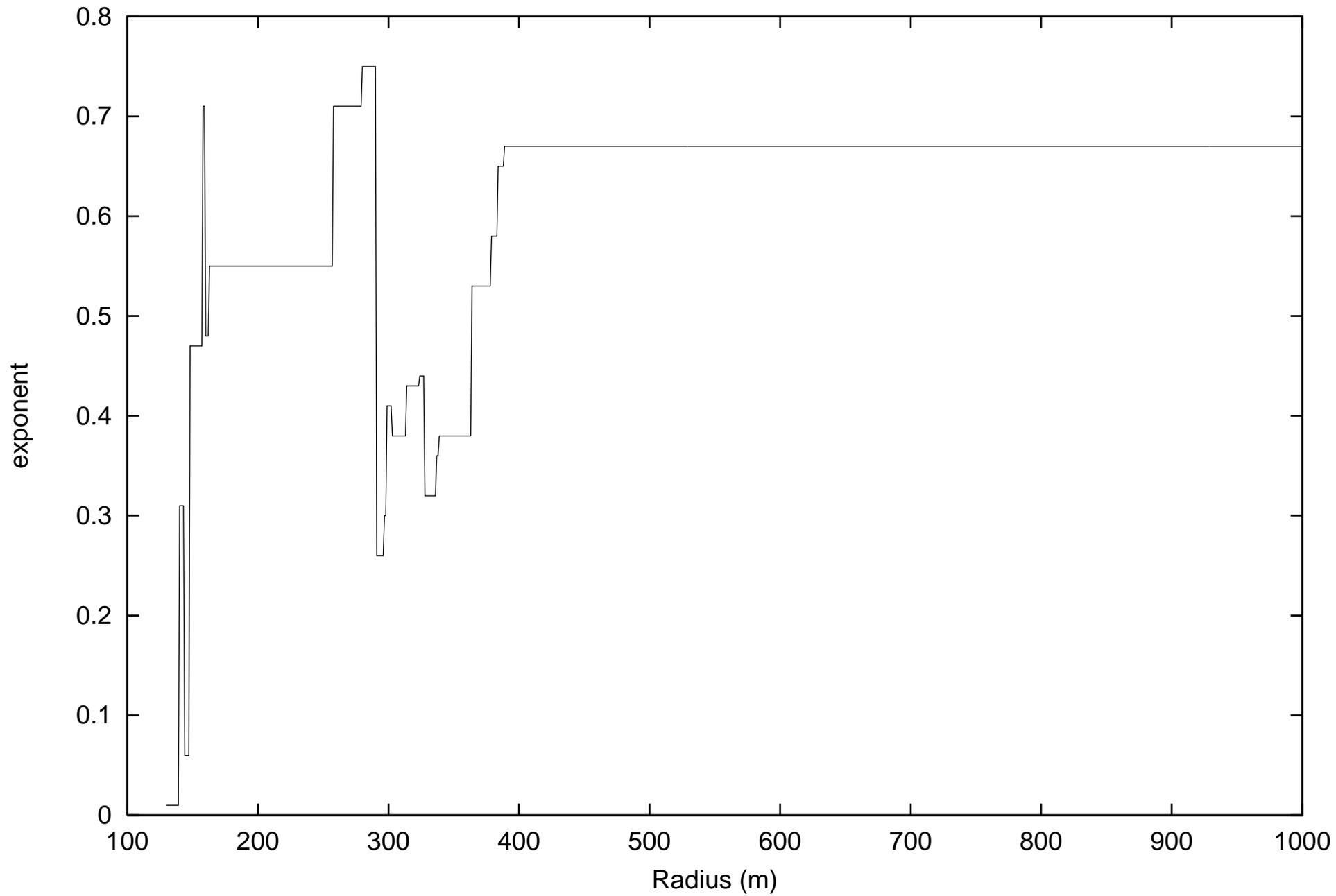

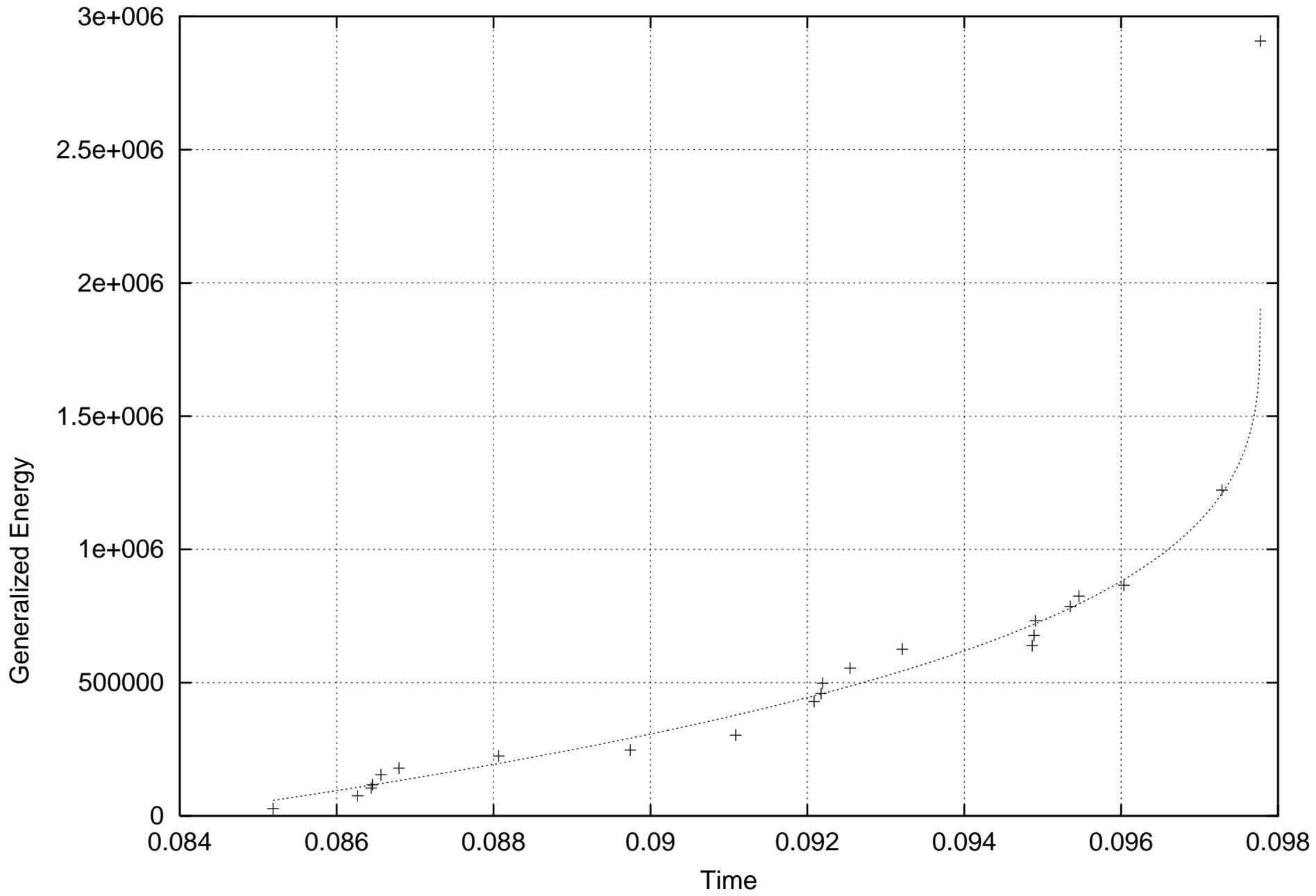

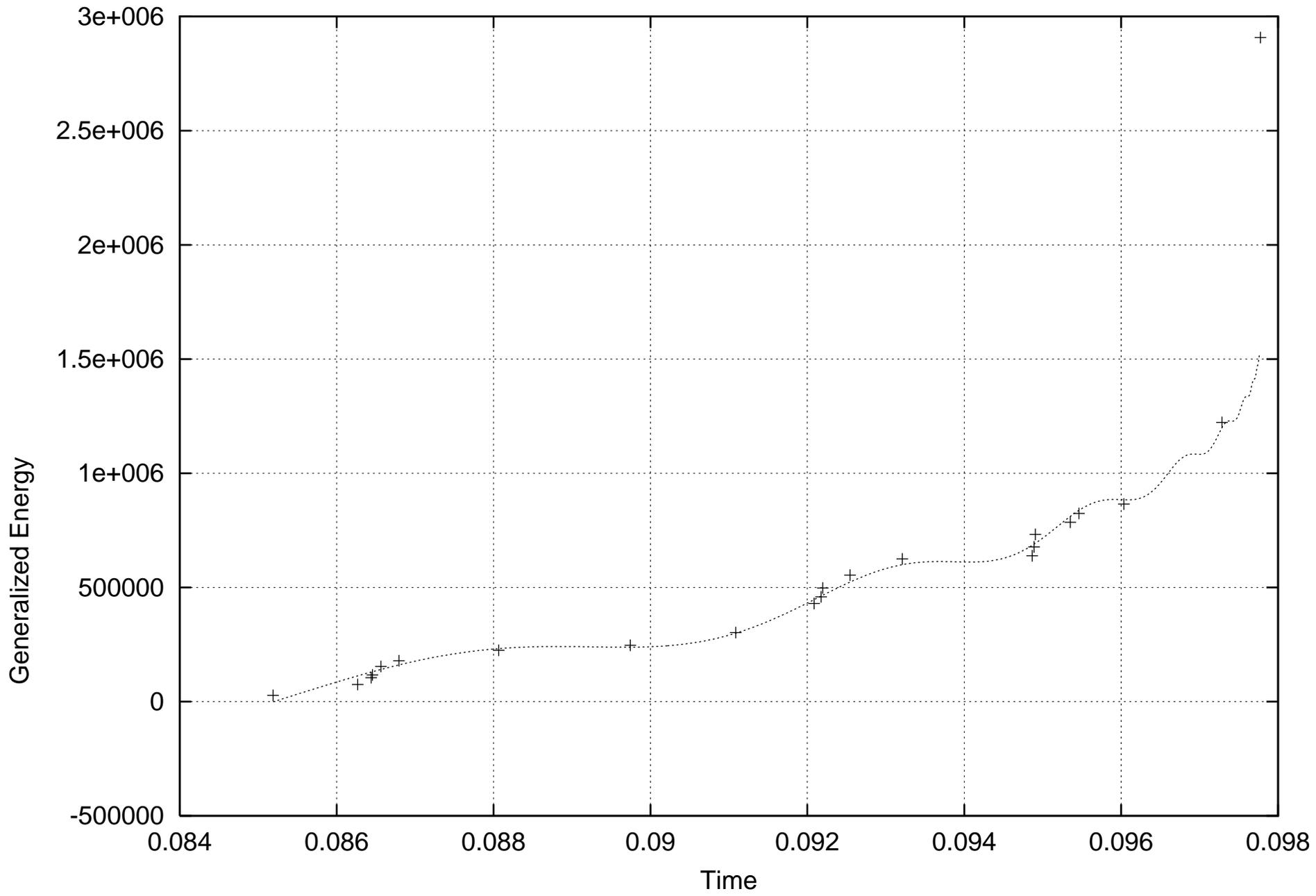

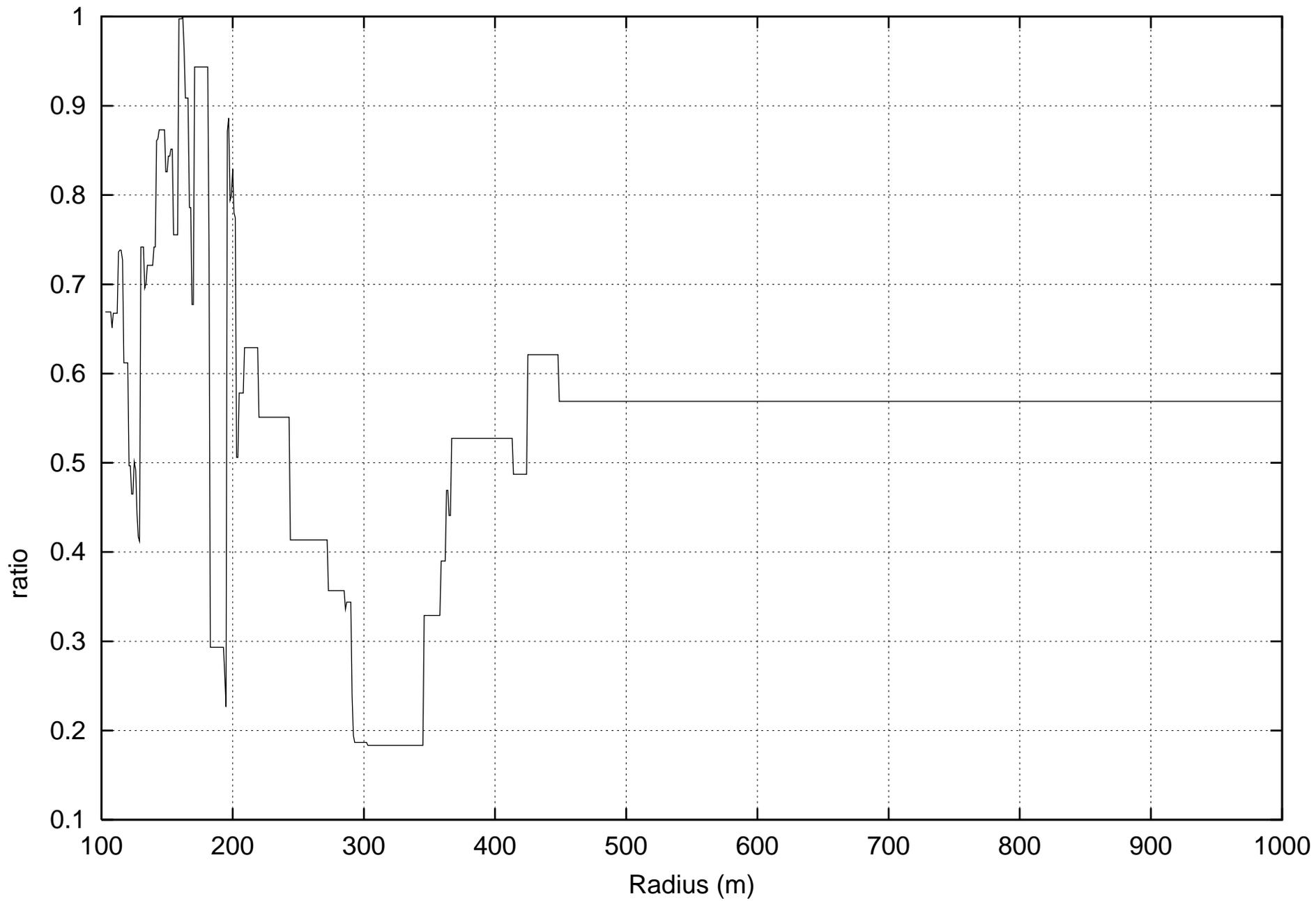

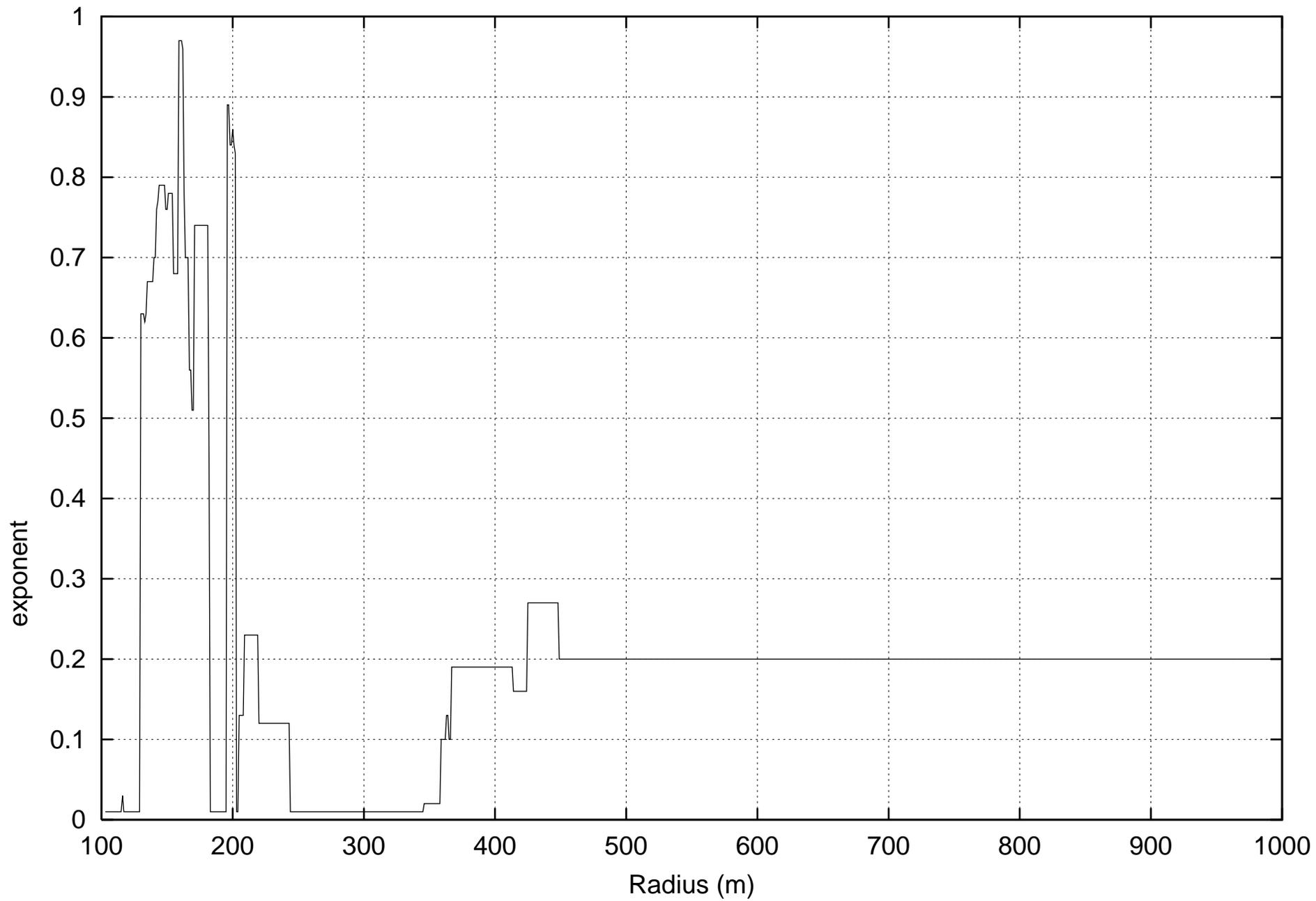

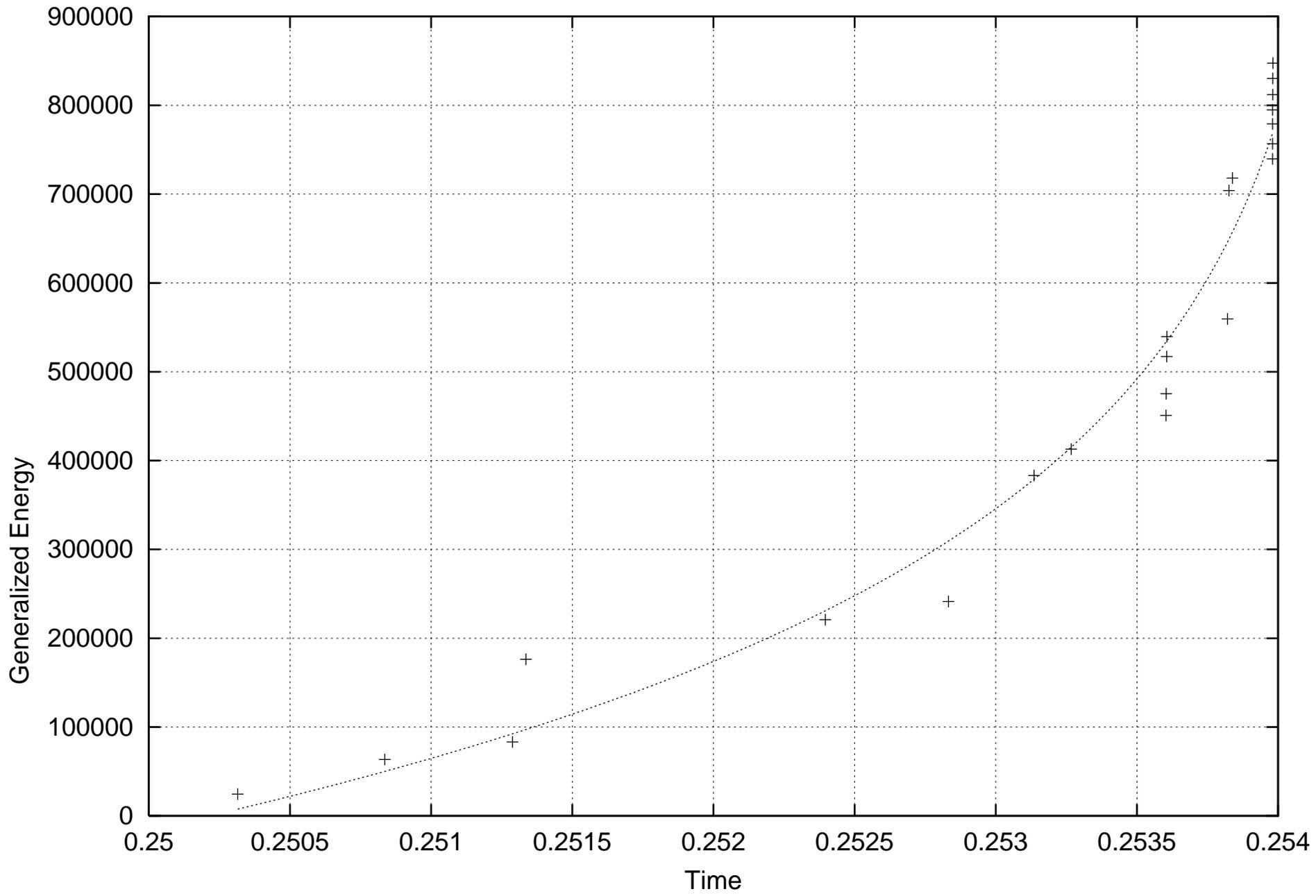

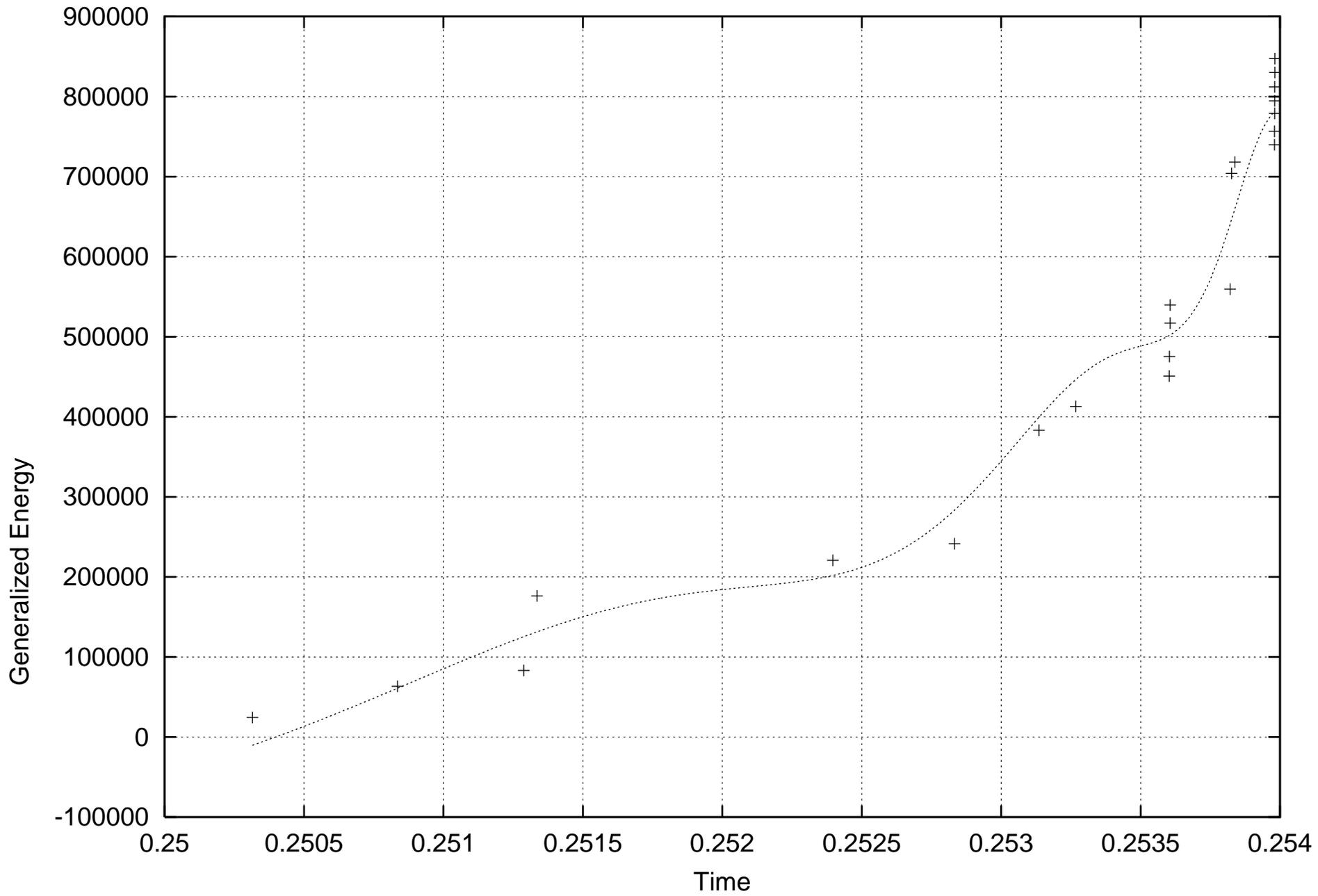

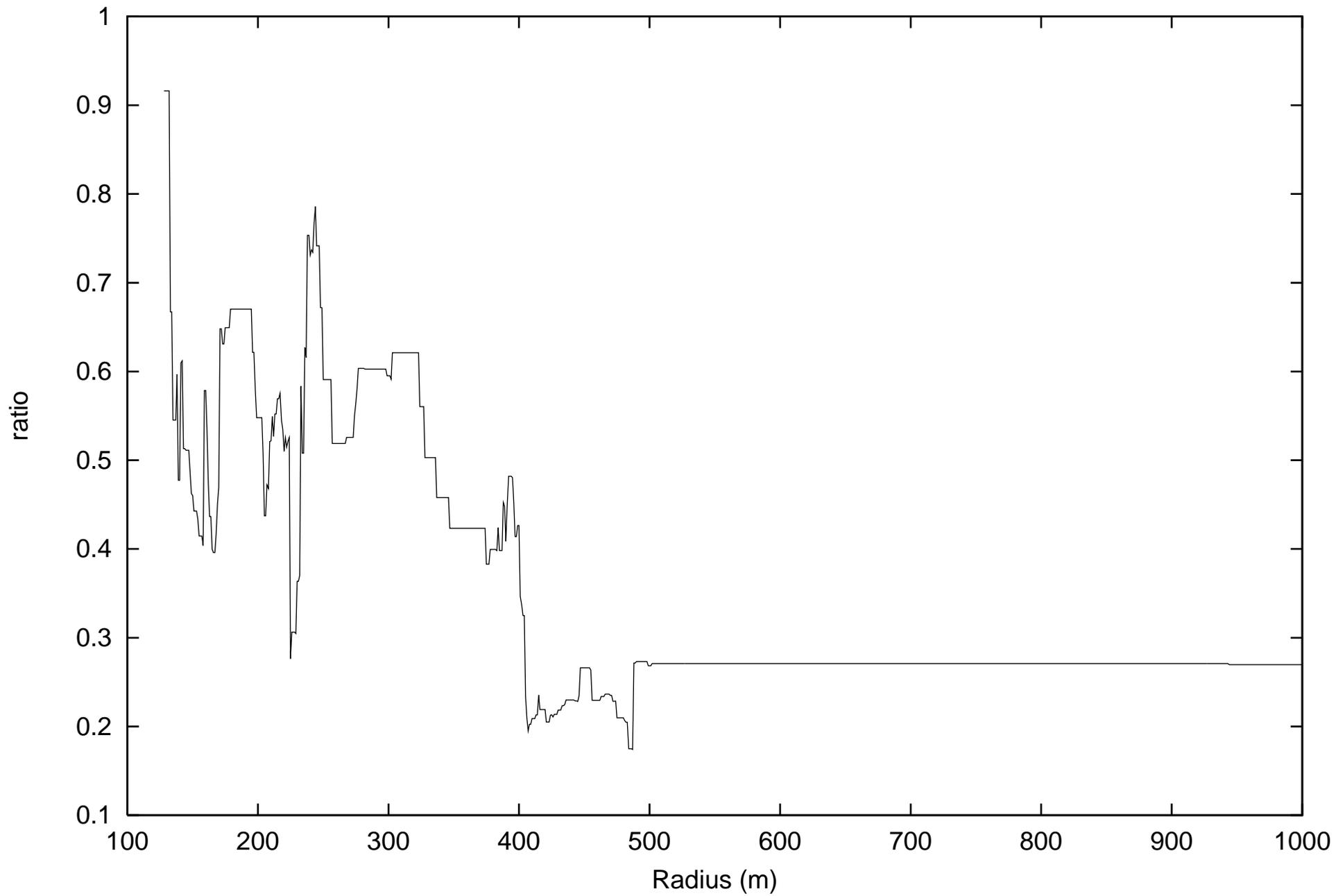

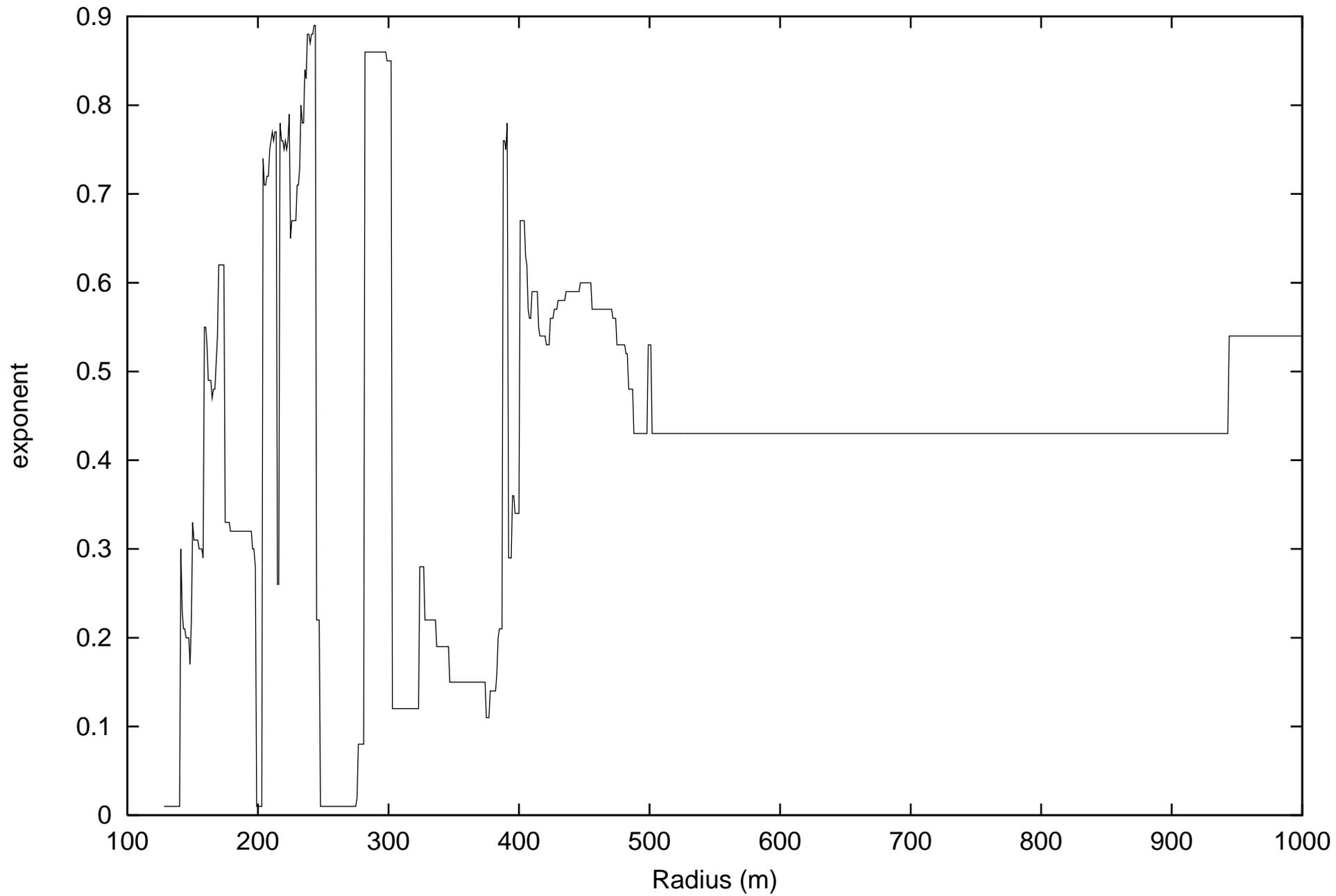

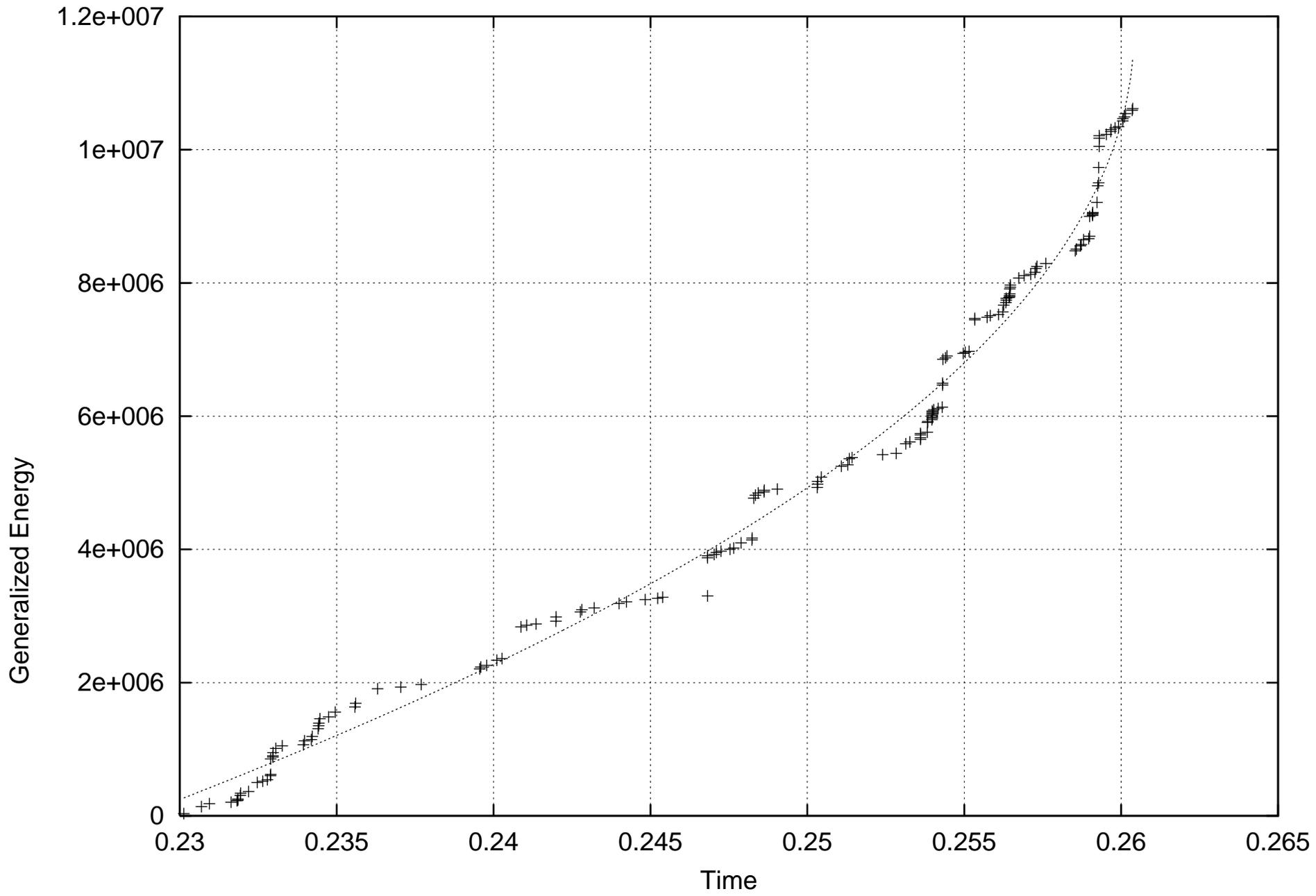

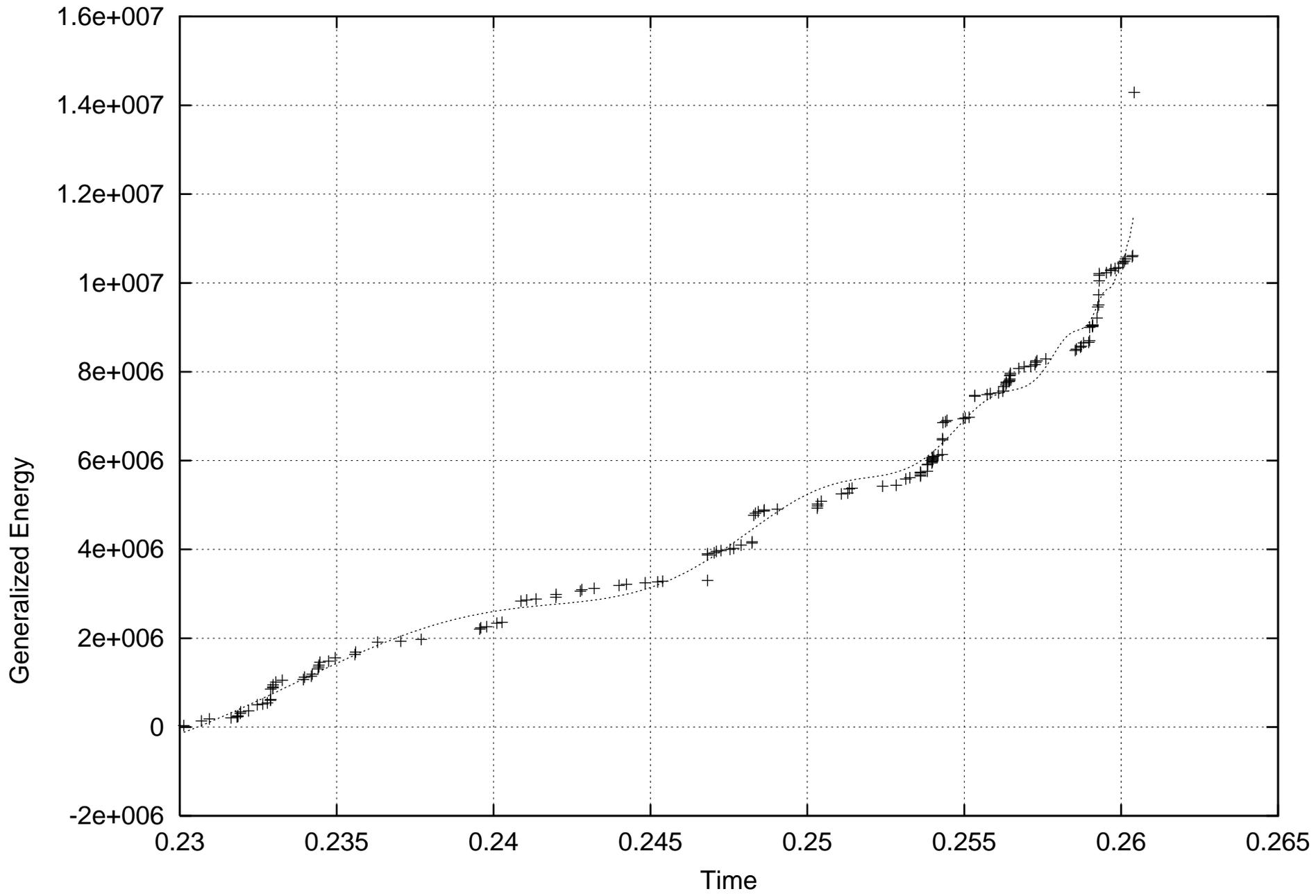

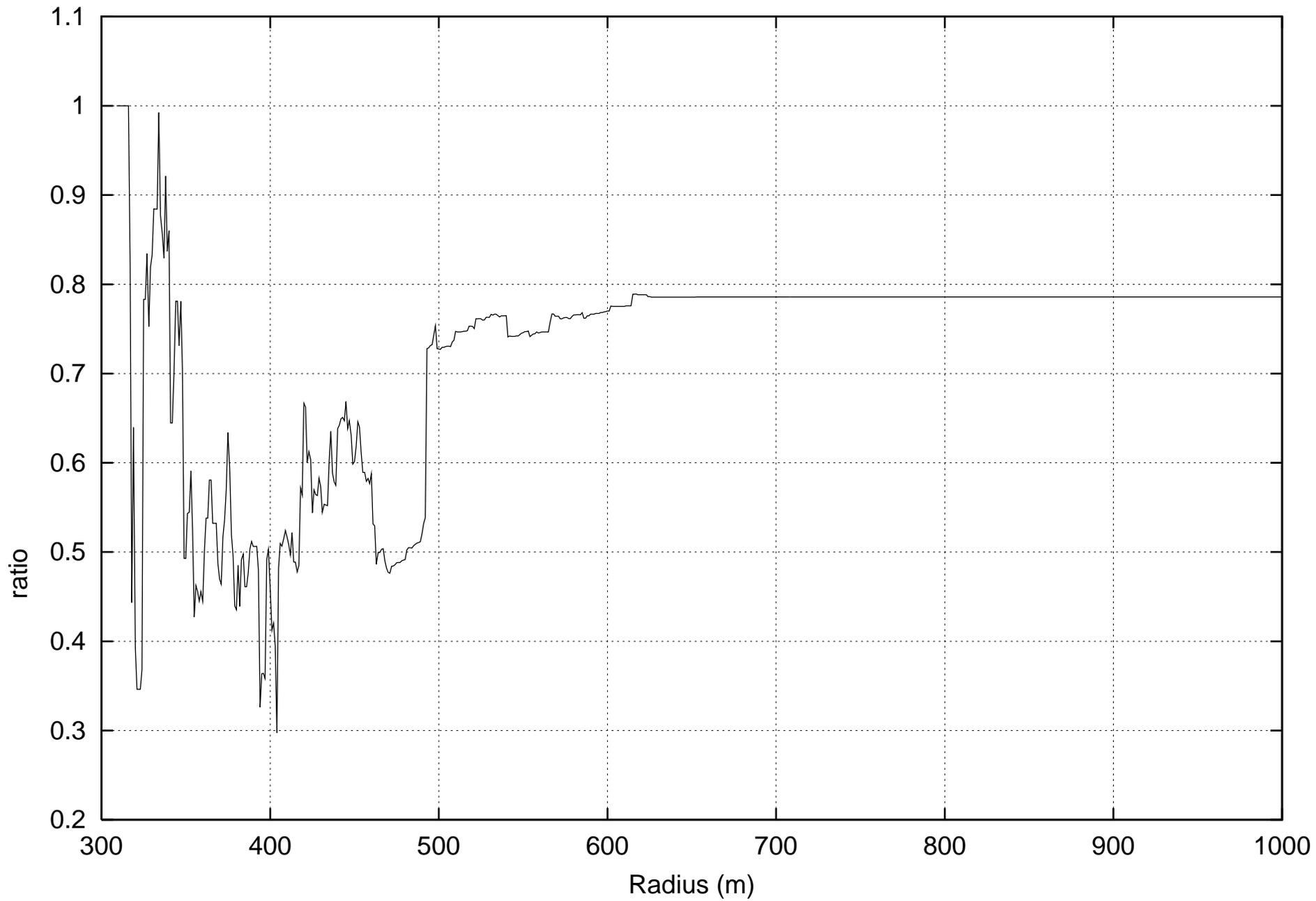

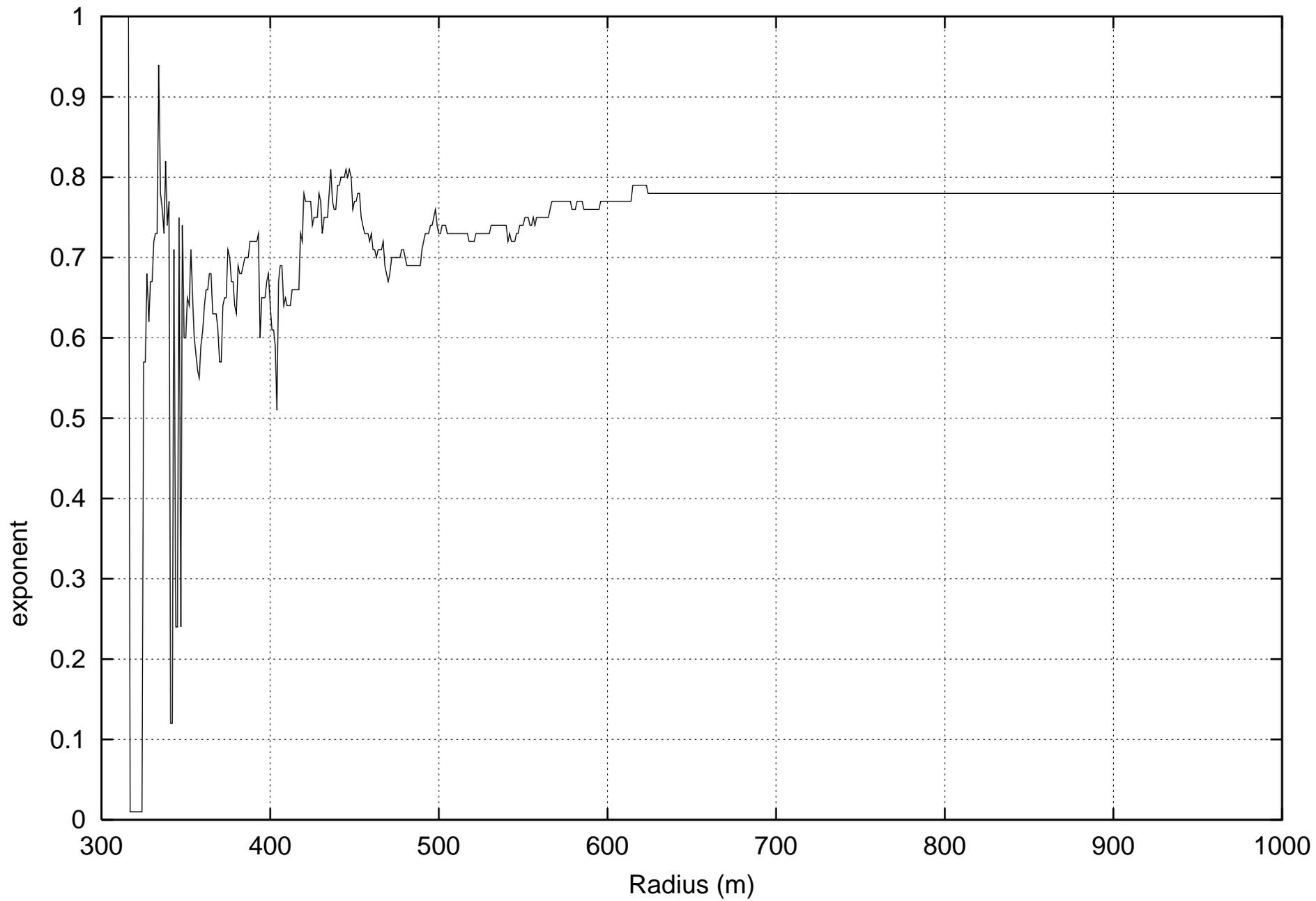

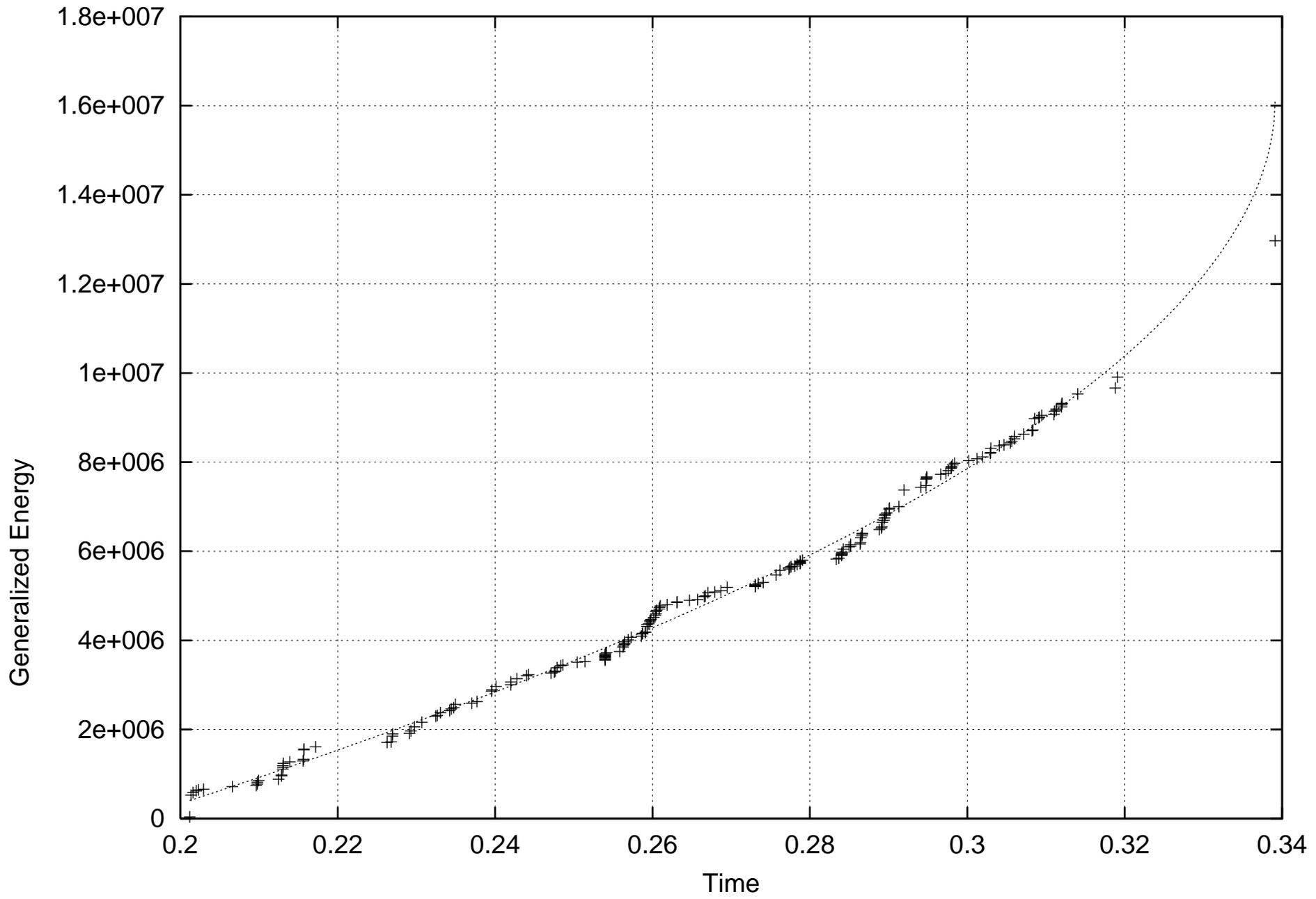

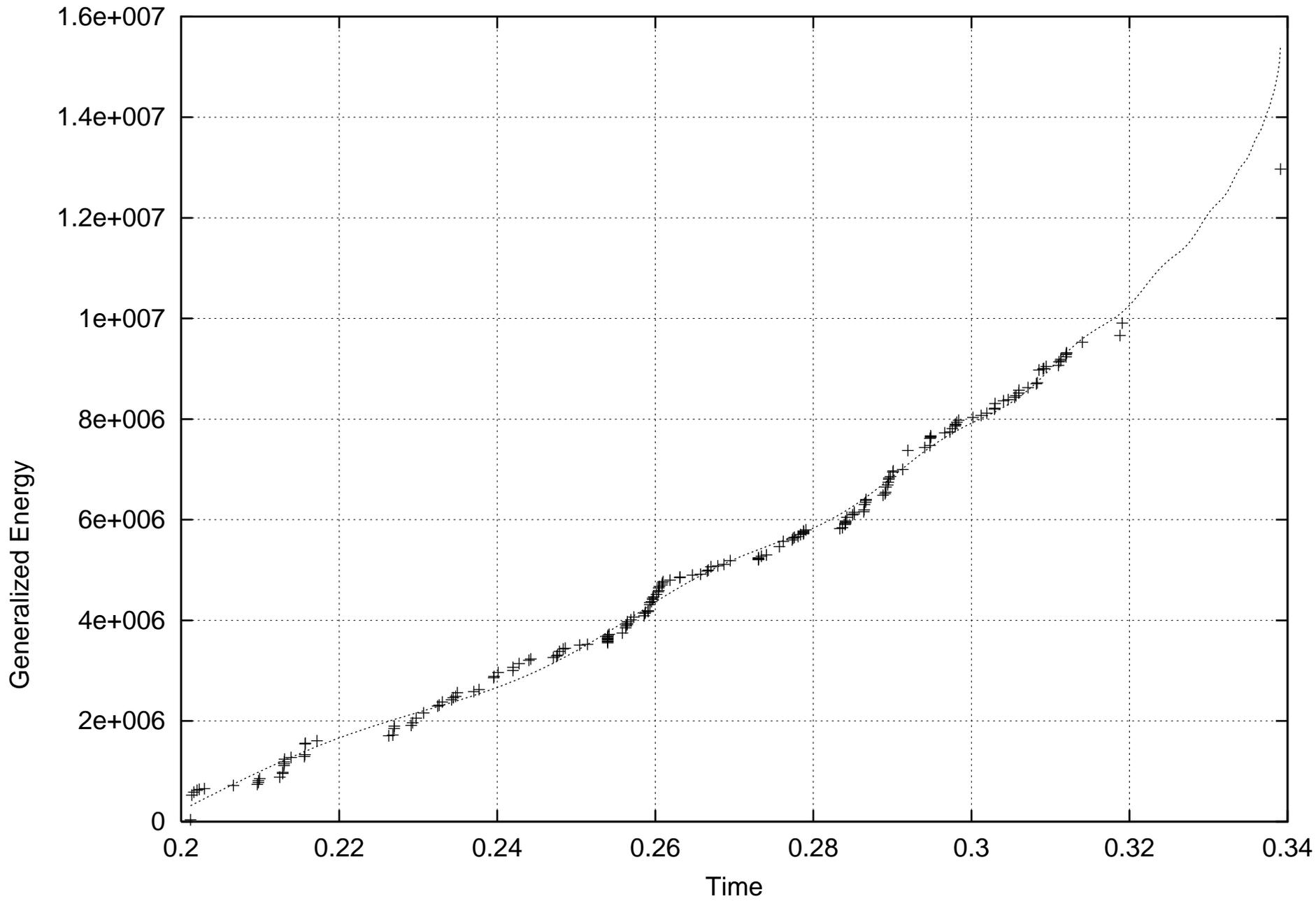

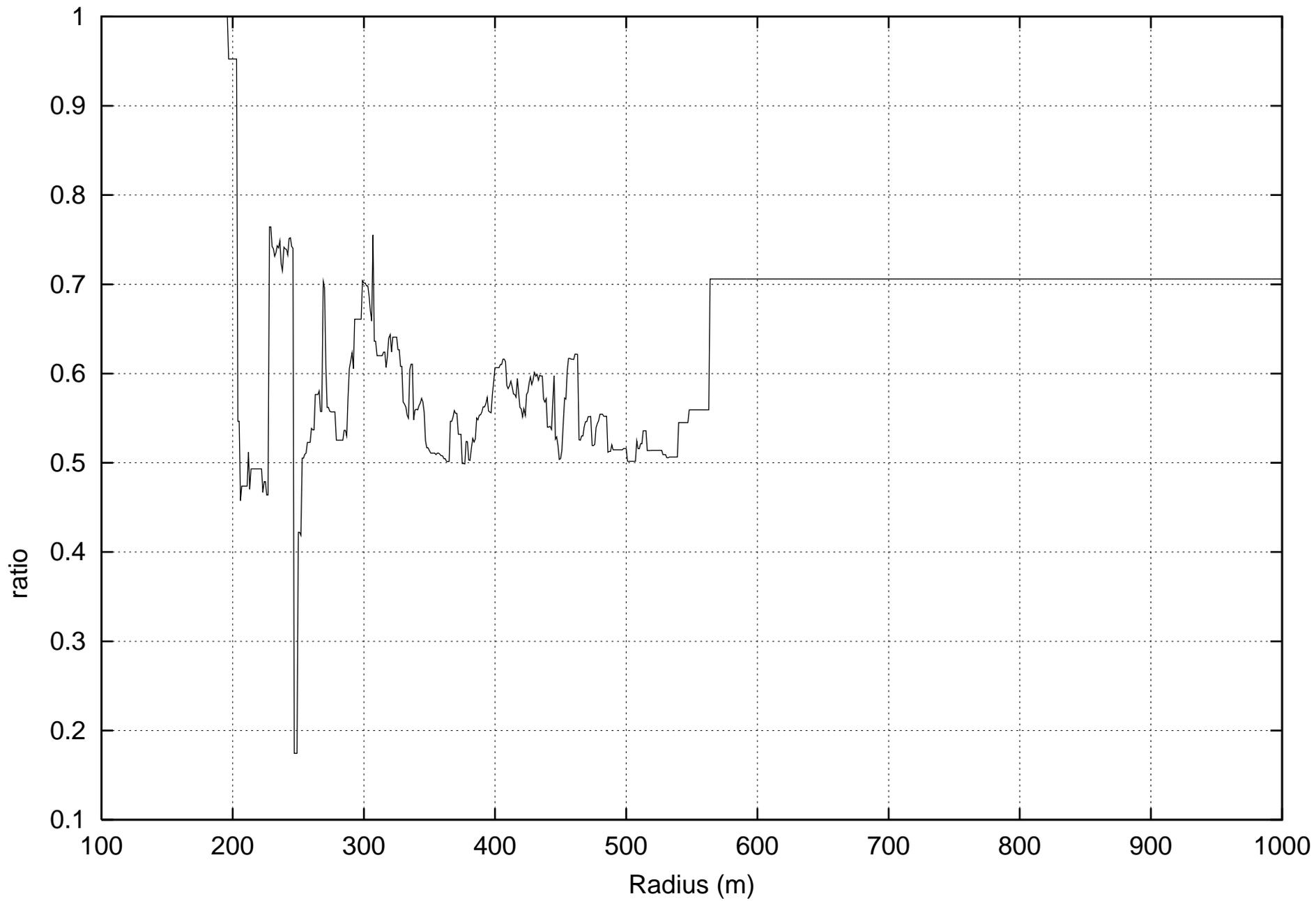

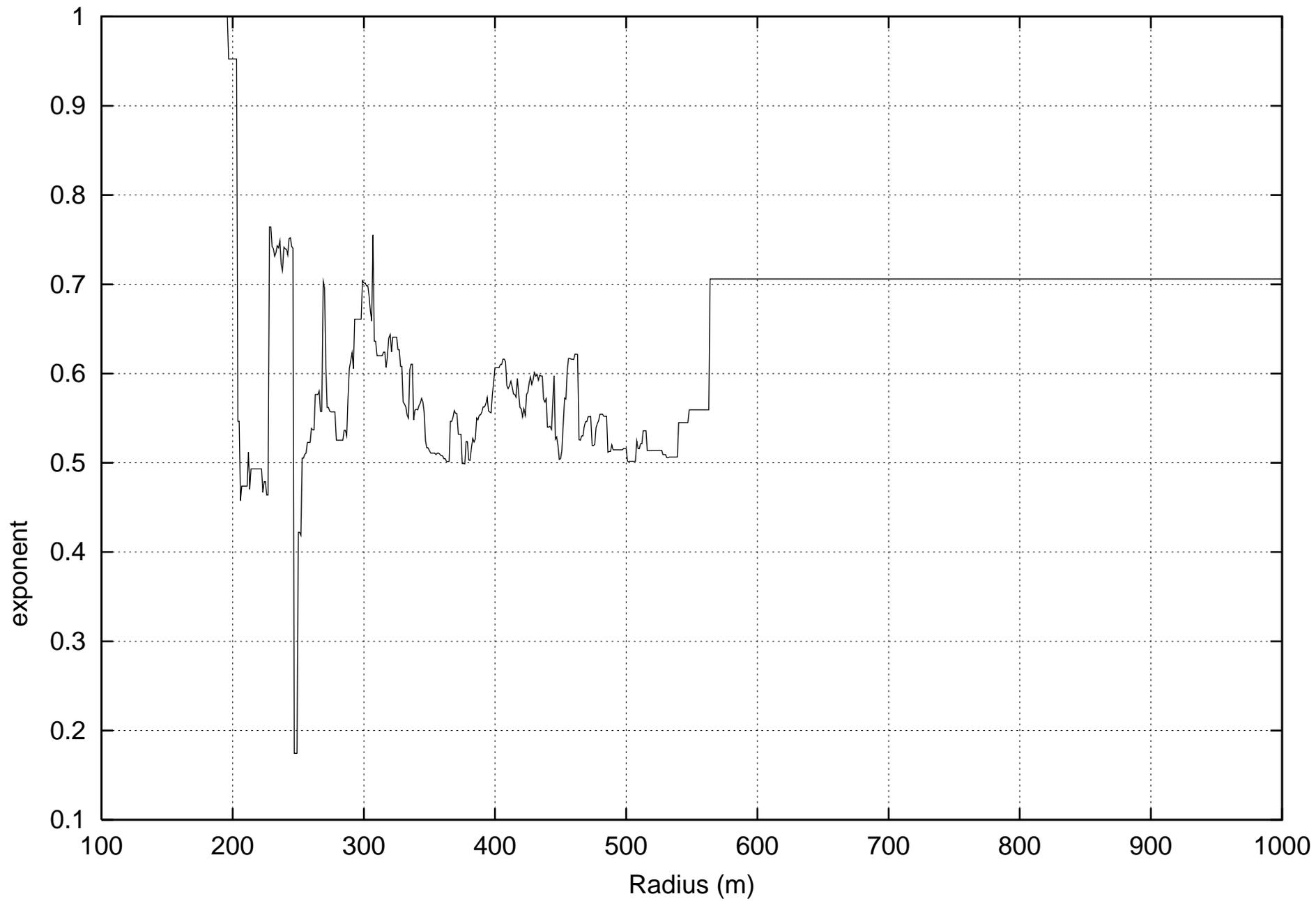

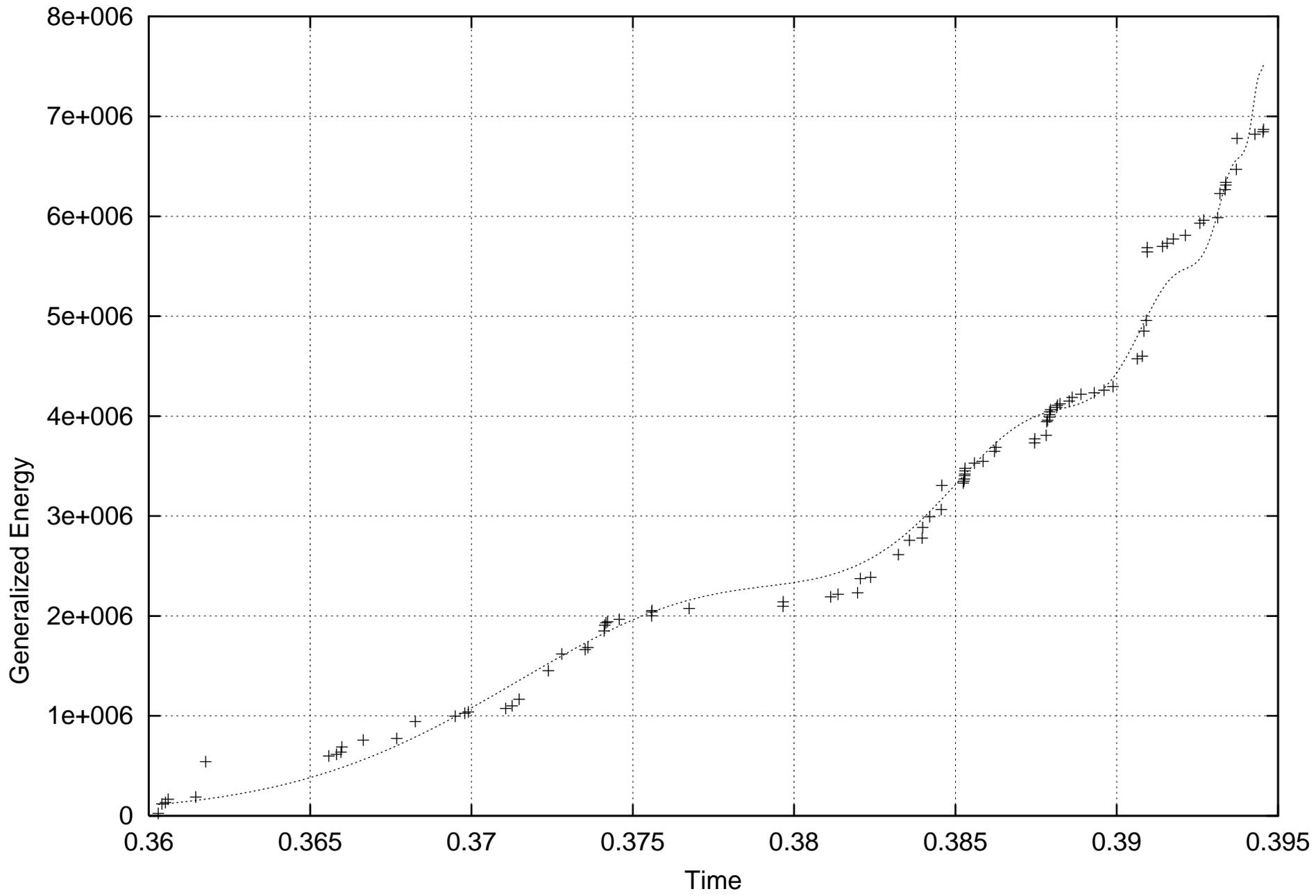

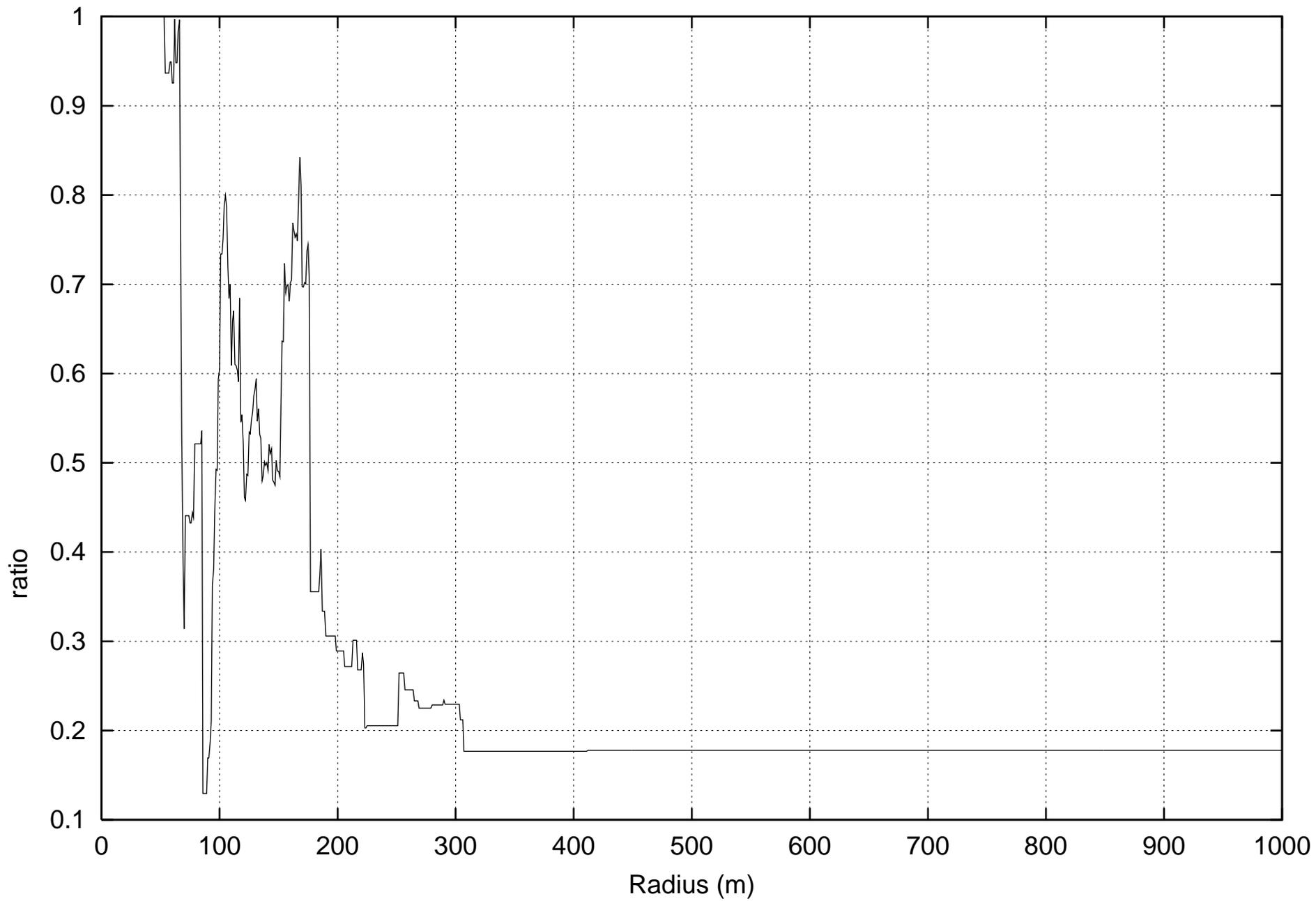

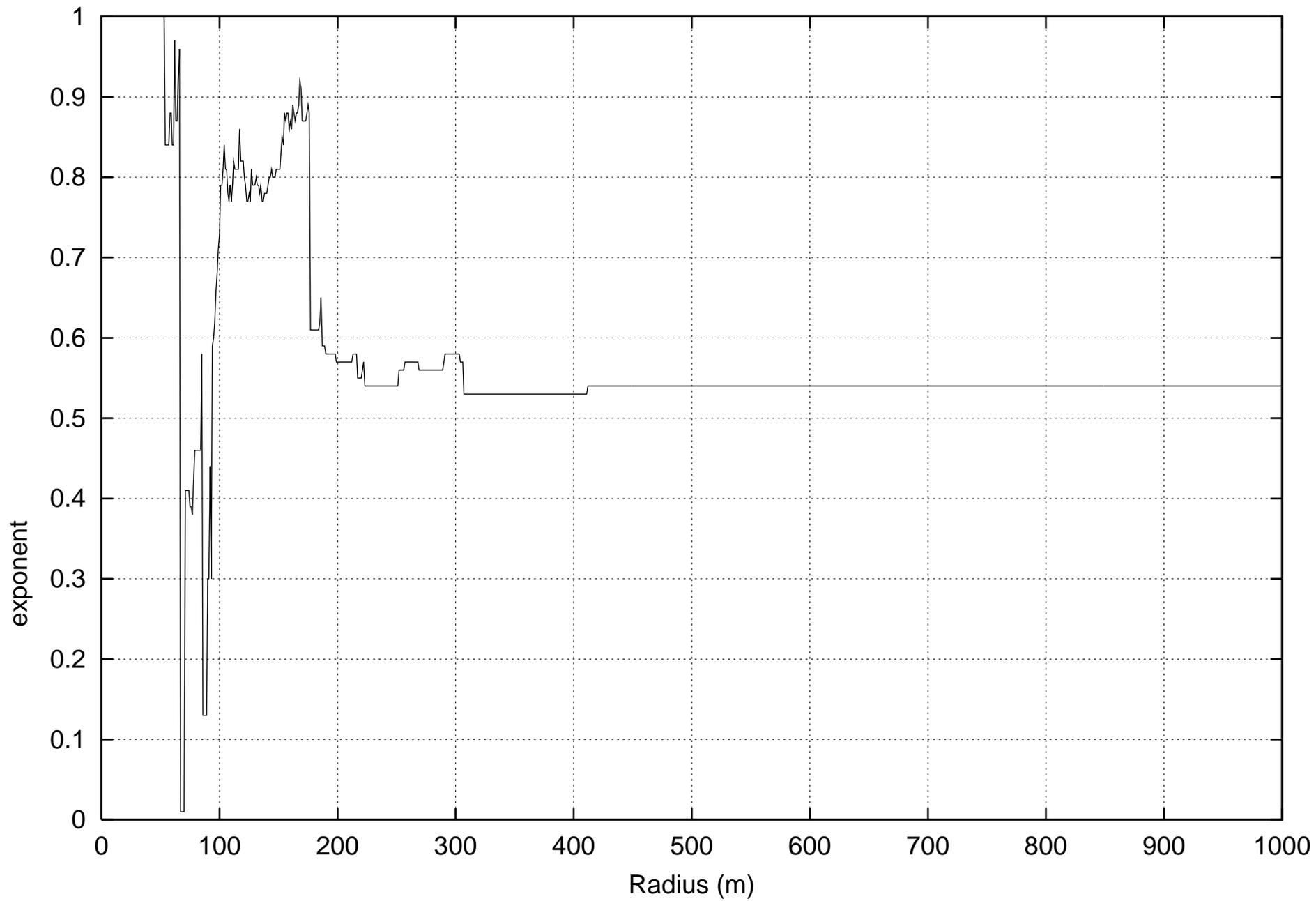

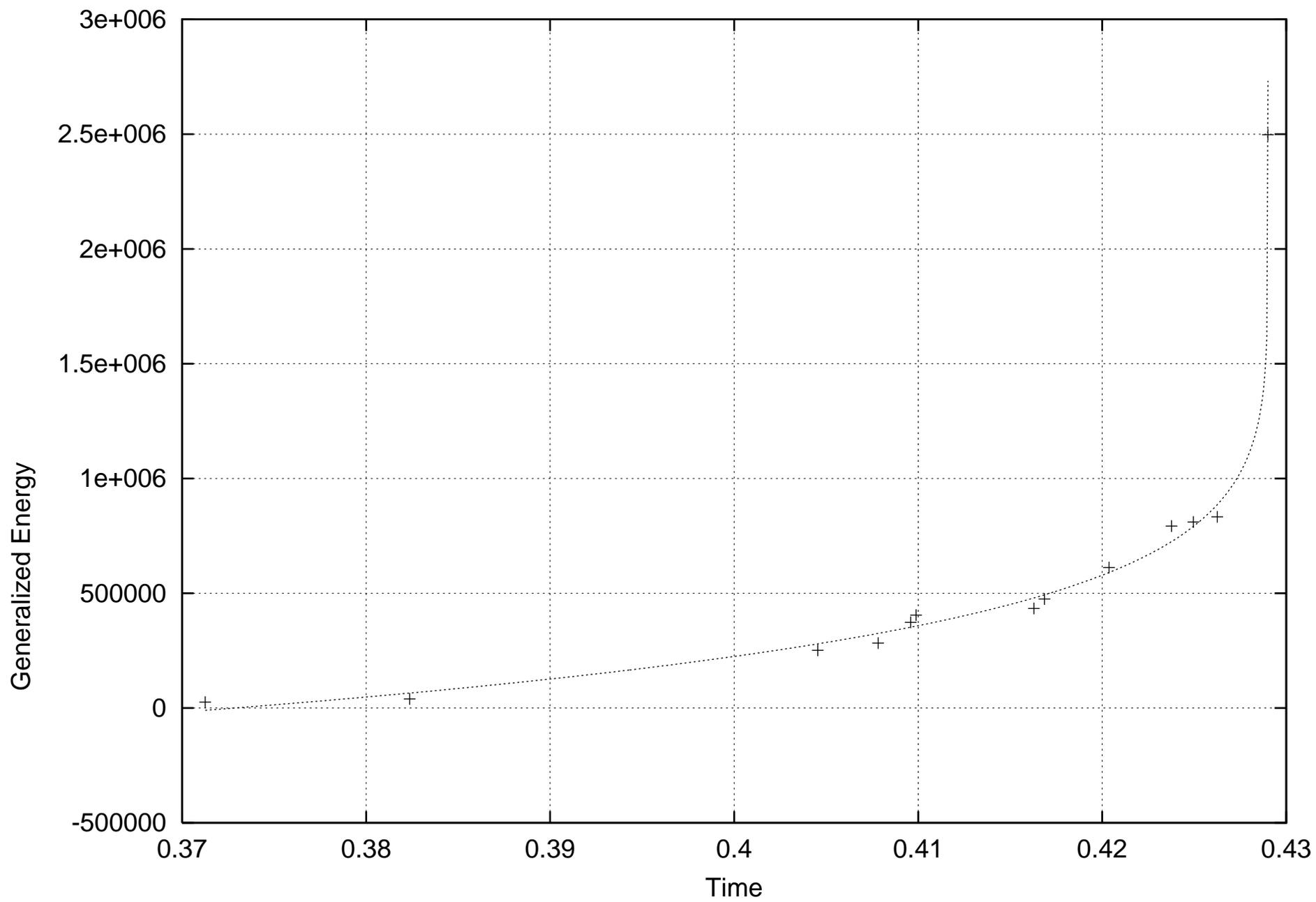

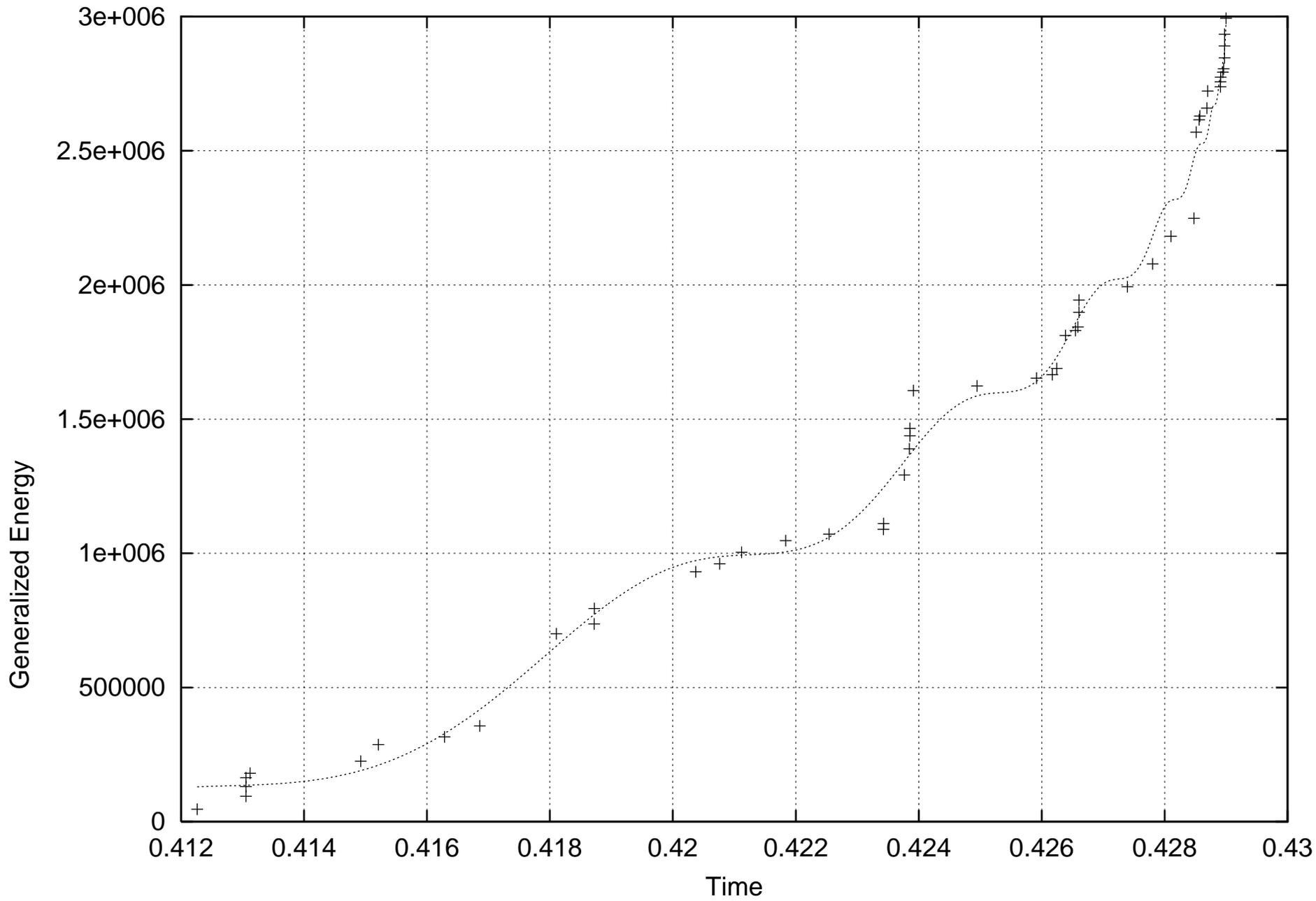

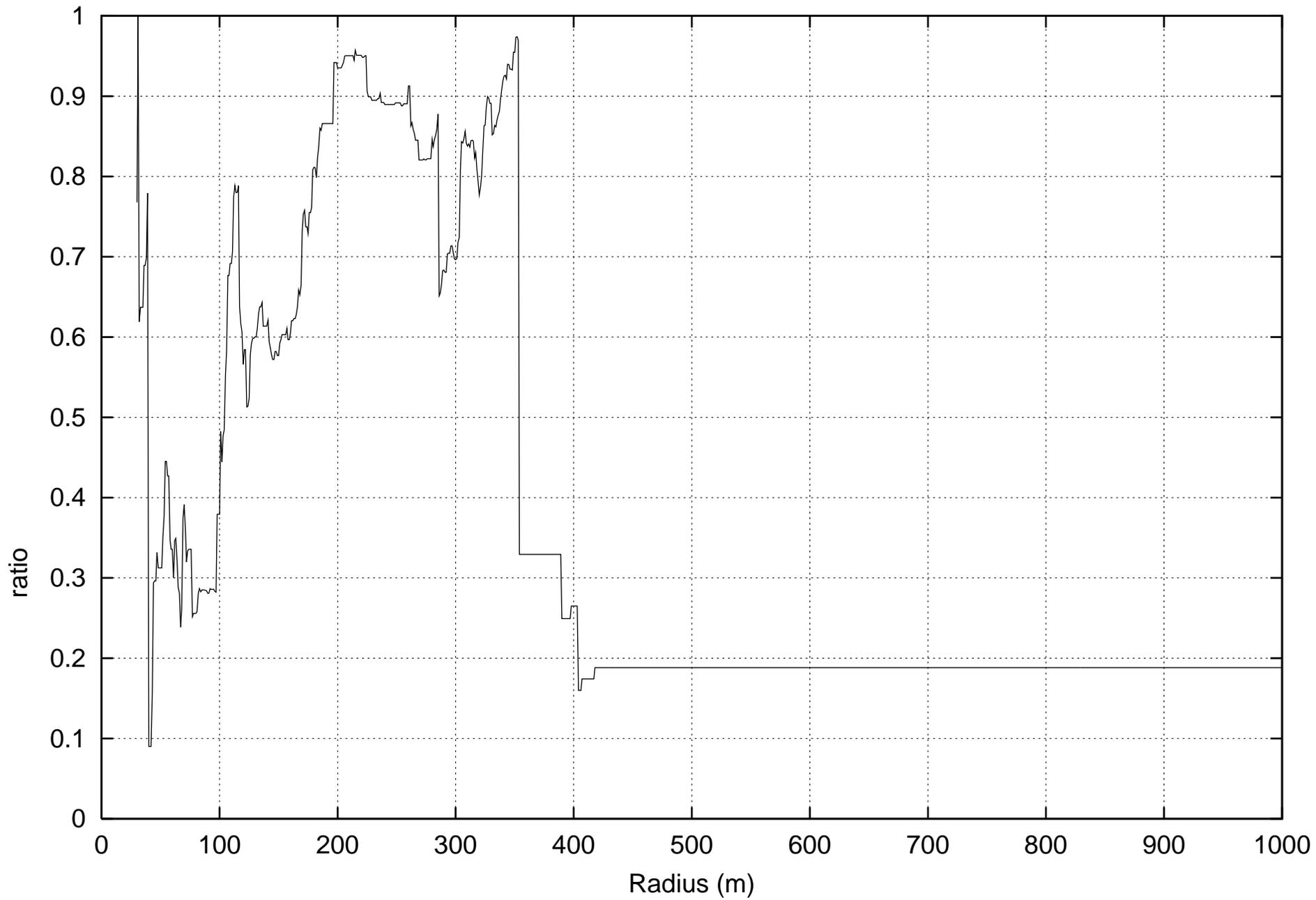

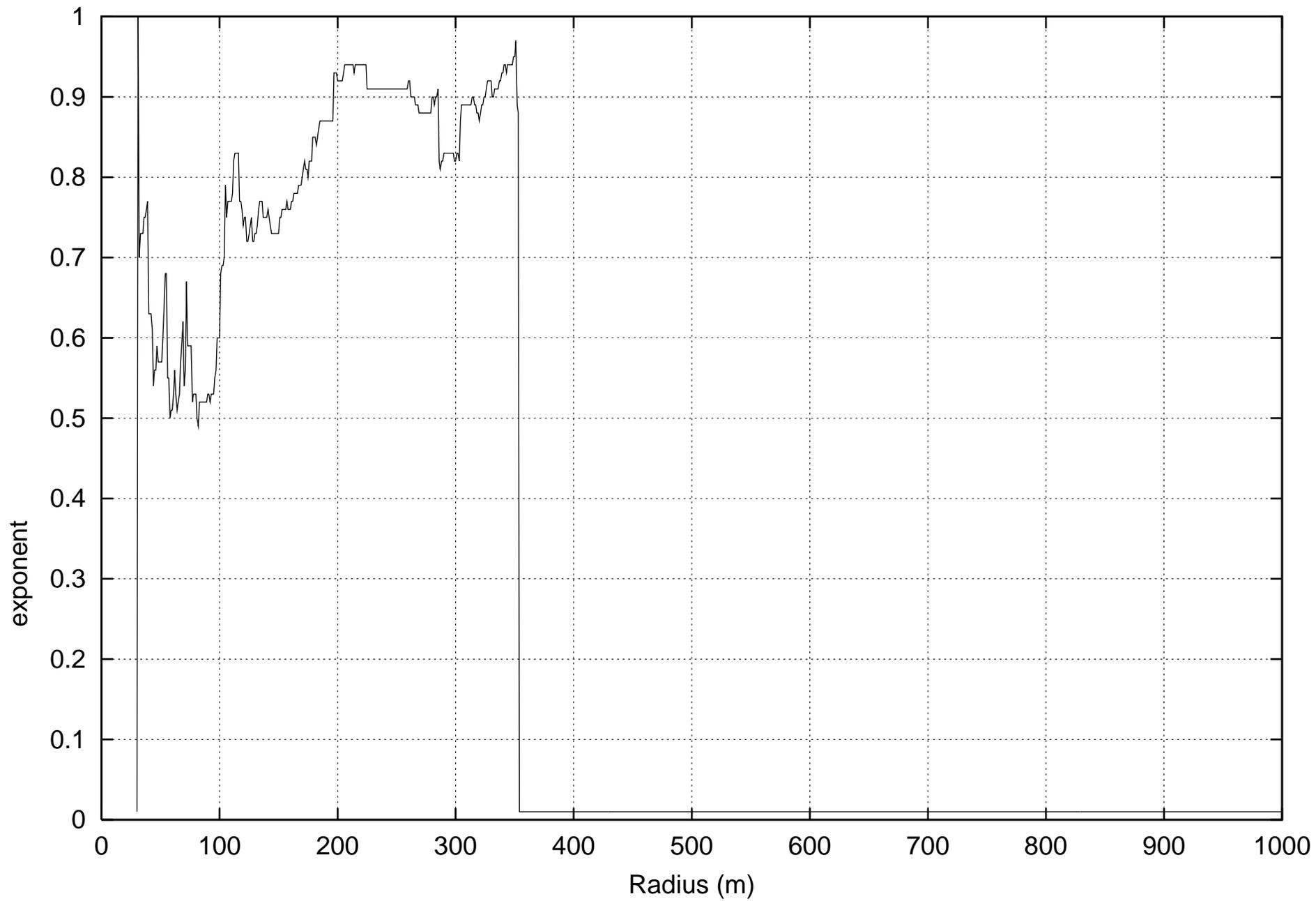

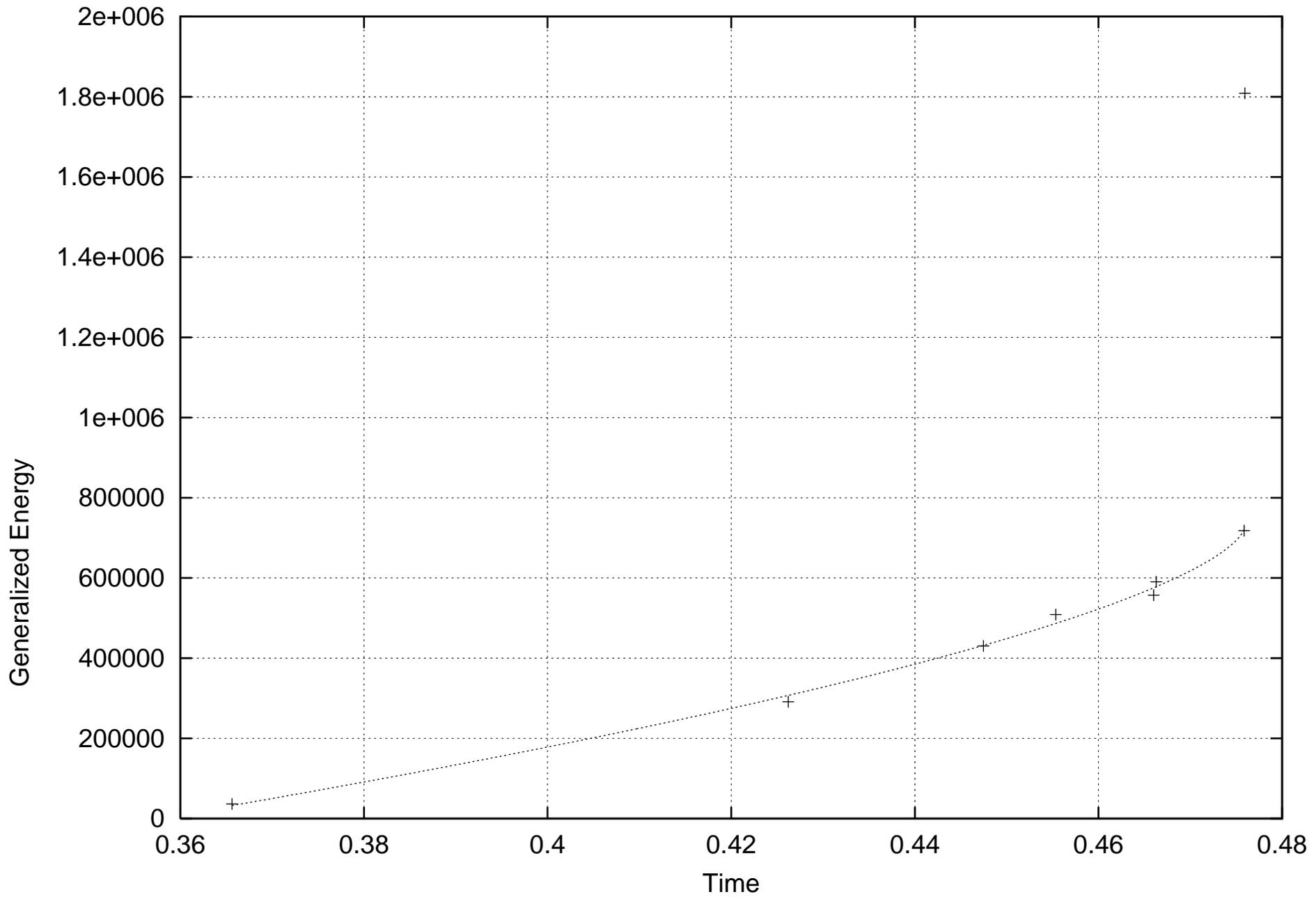

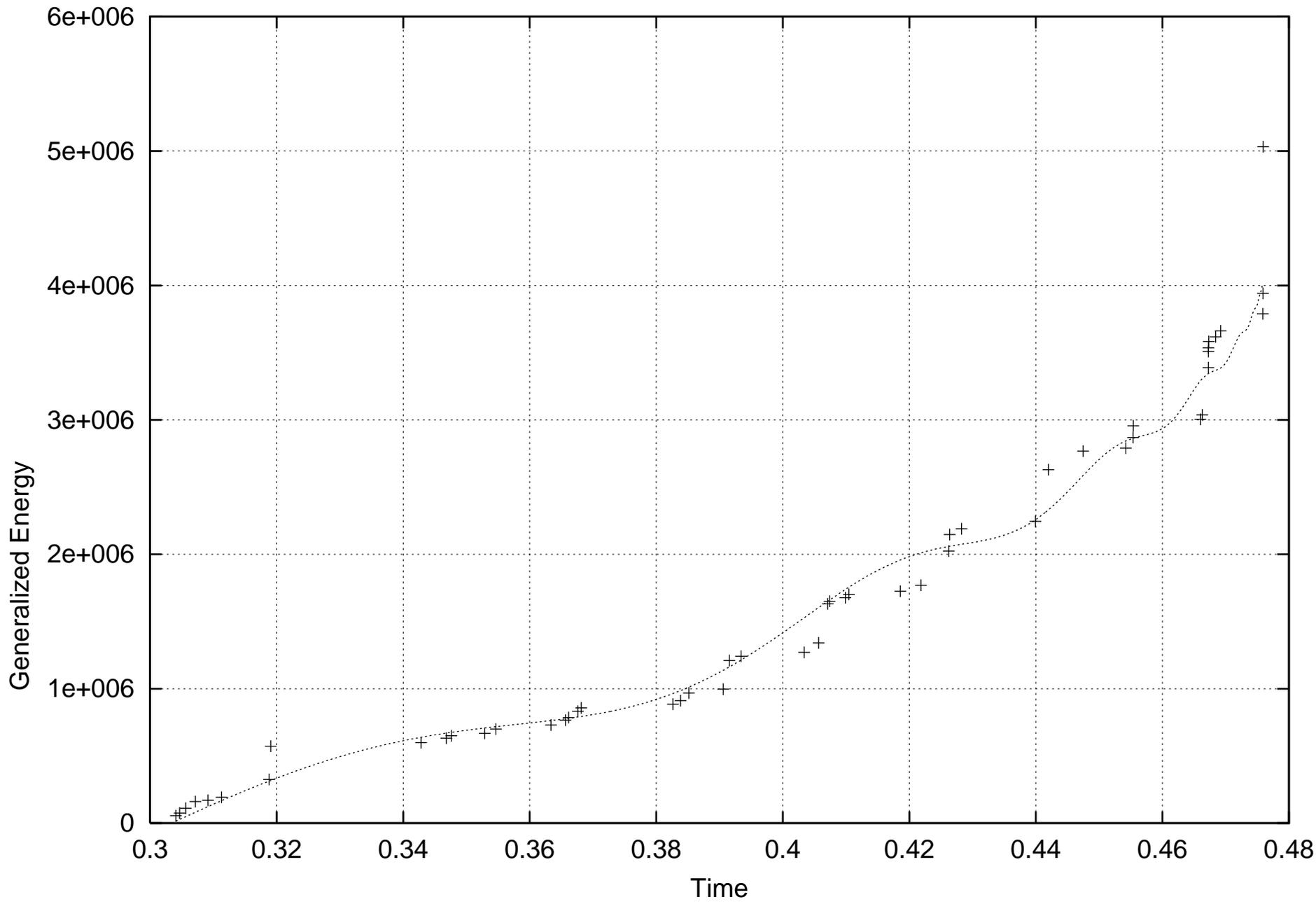

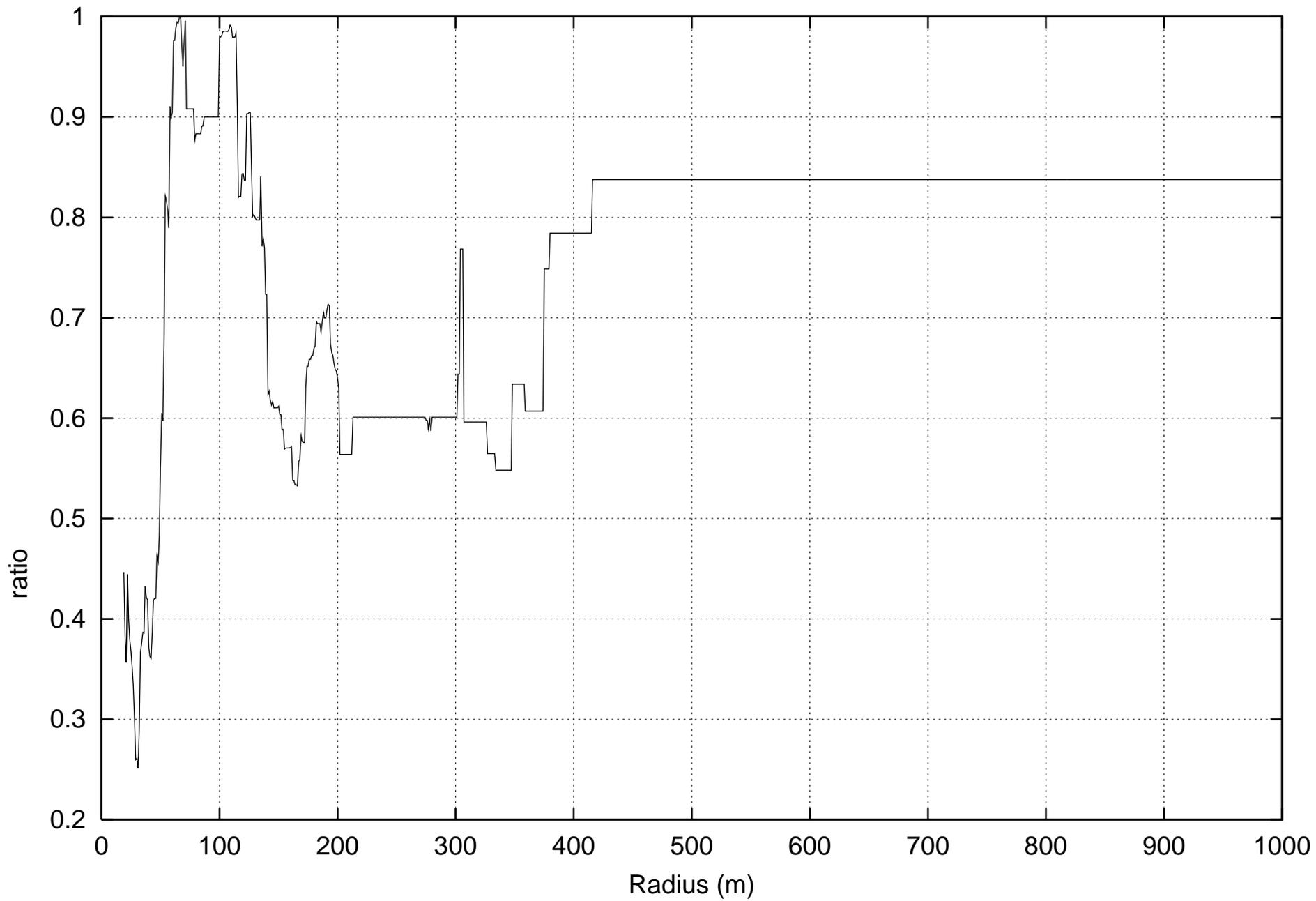

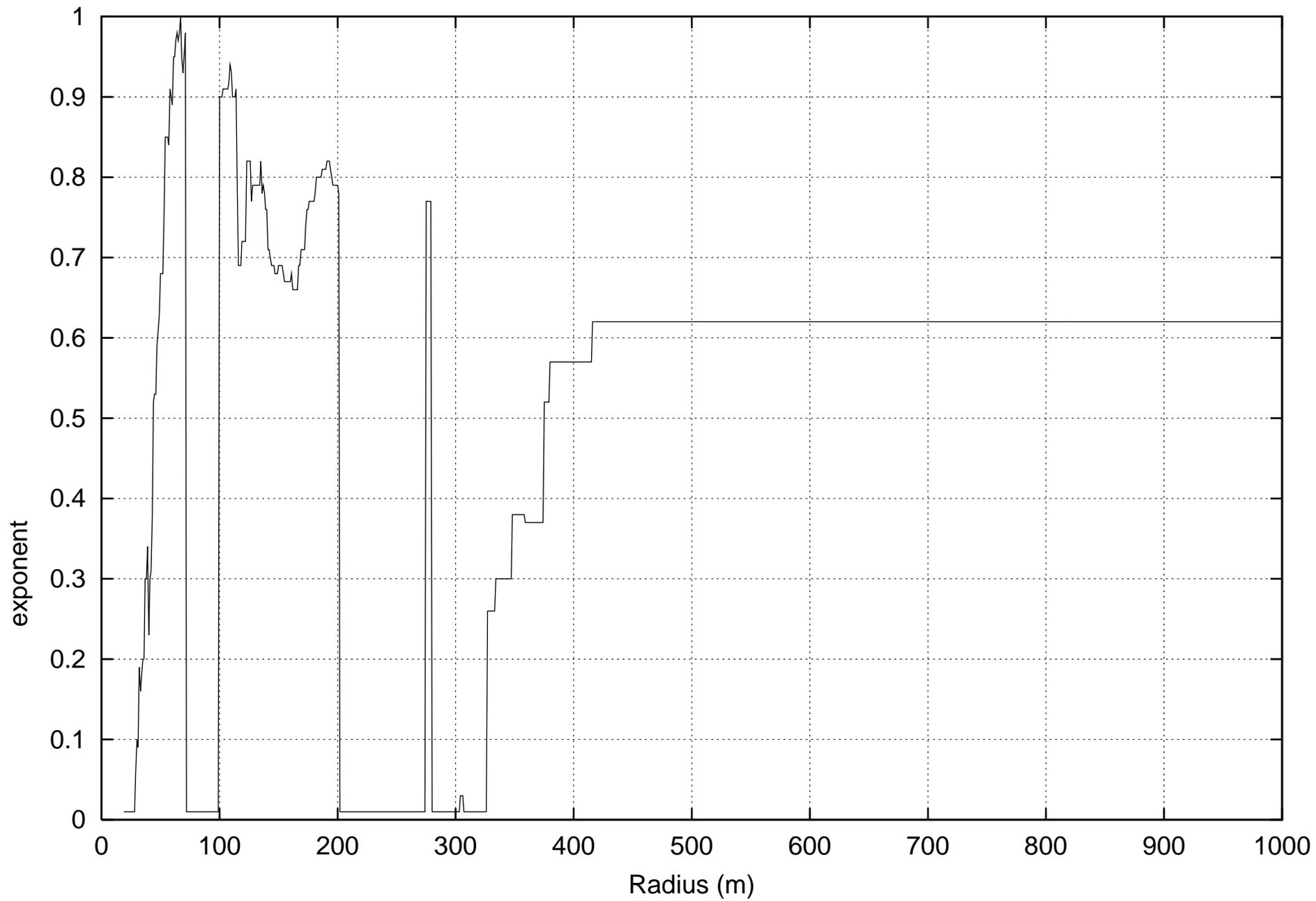

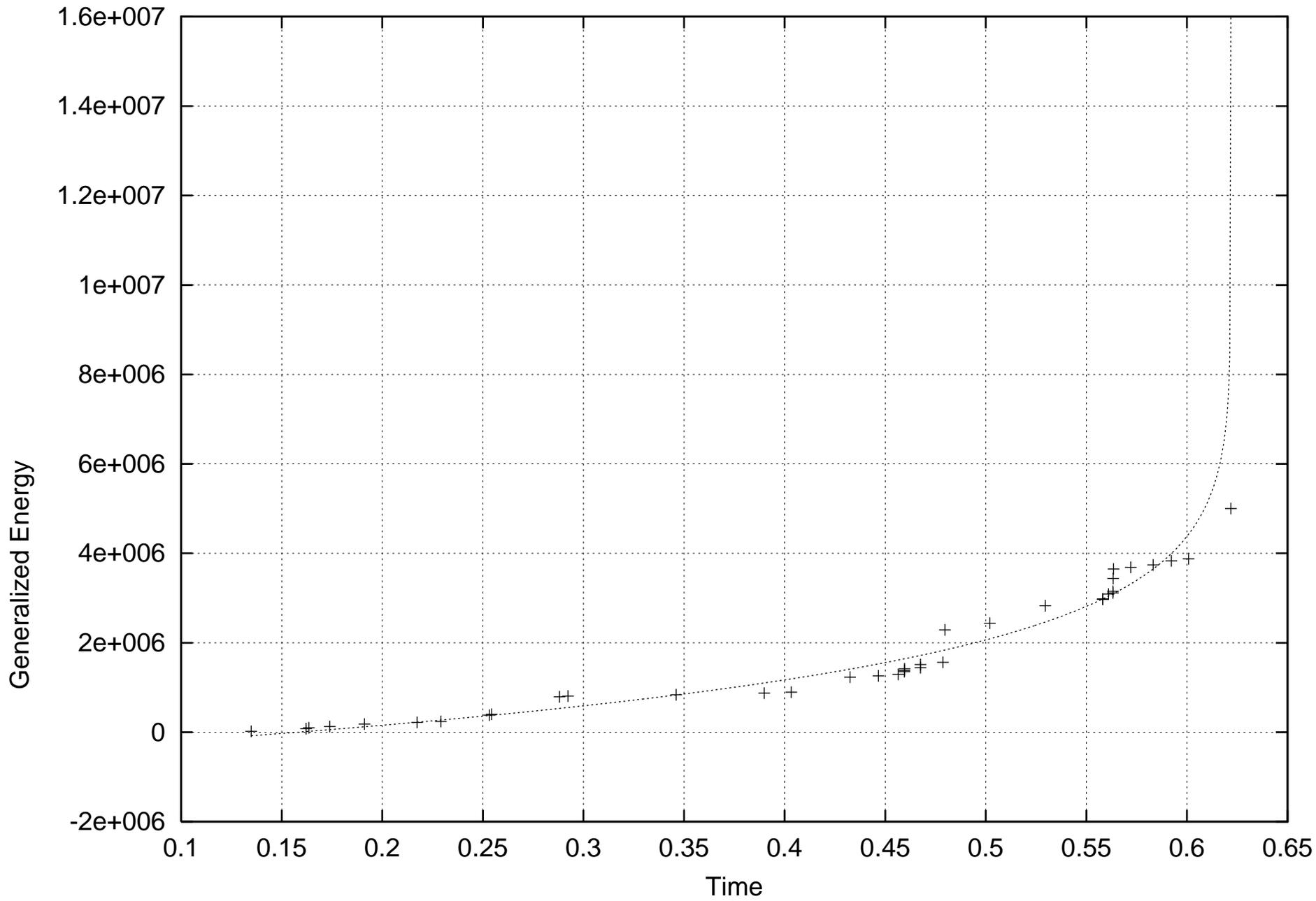

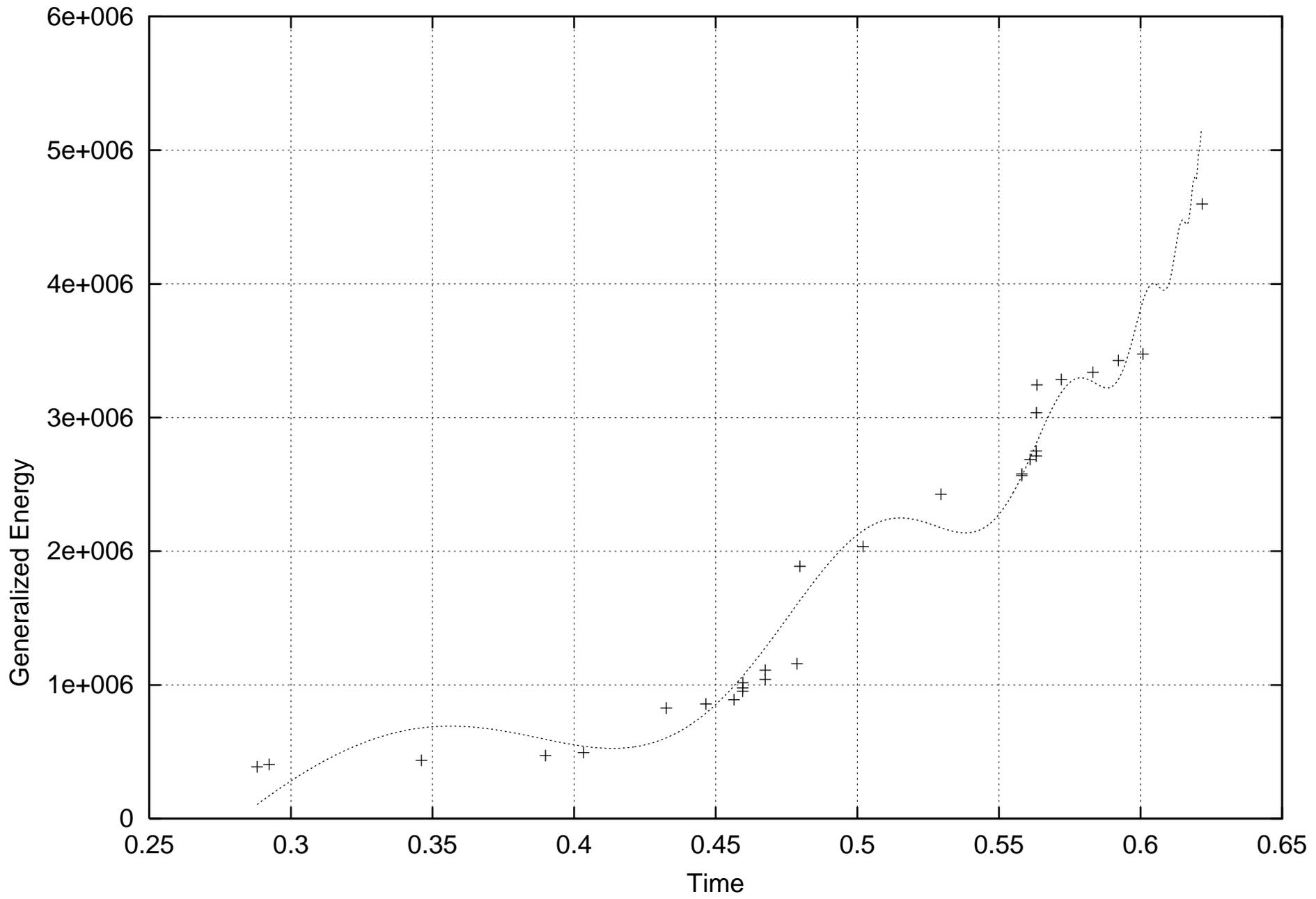

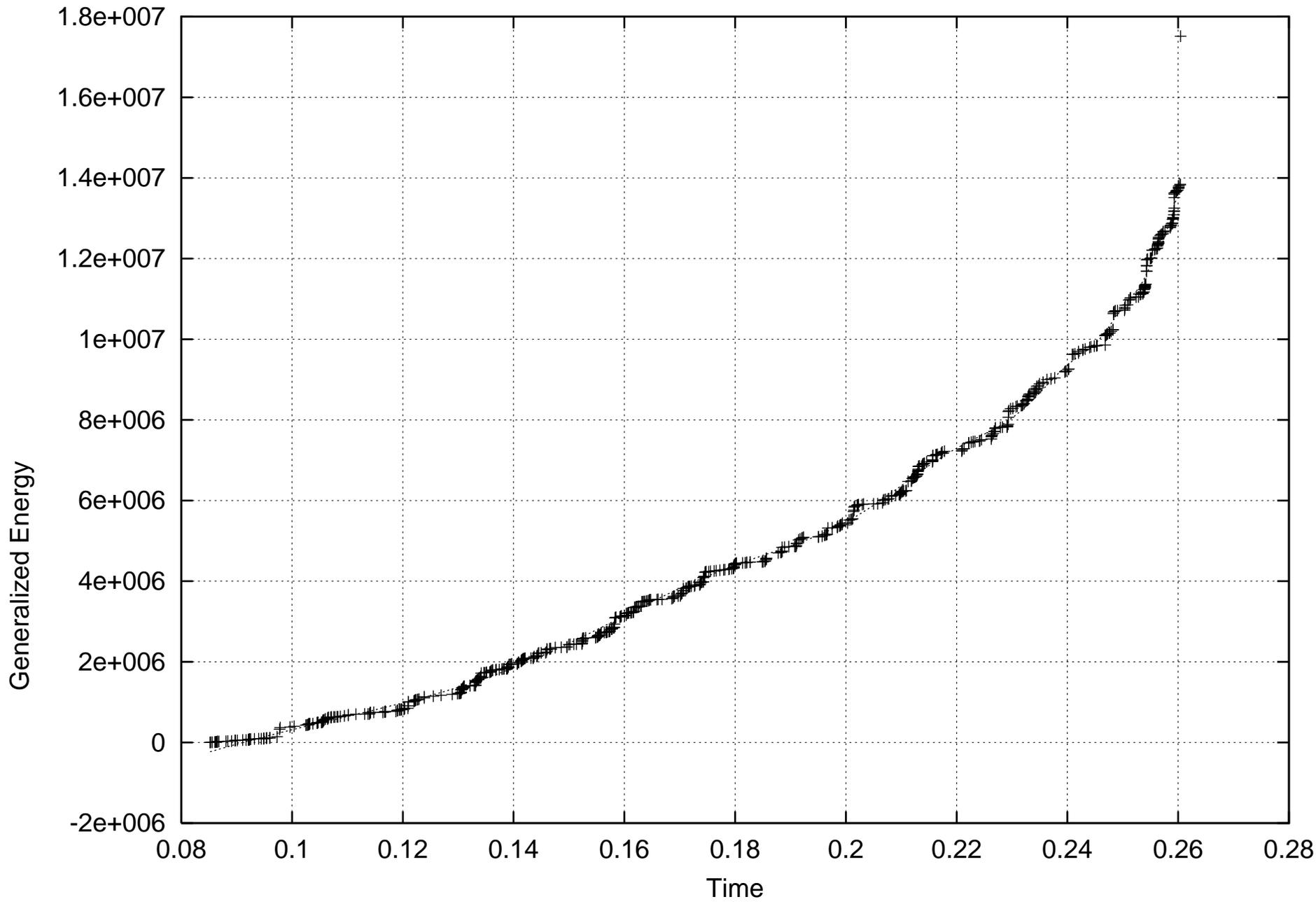

# The critical earthquake concept applied to mine rockbursts with time-to-failure analysis

**G. Ouillon[1] and D. Sornette[1,2]**

1-Laboratoire de Physique de la Matiere Condensee, CNRS UMR6622
Universite de Nice-Sophia Antipolis, Faculte des Sciences, B.P. 71
06108 NICE Cedex 2, France
2-Institute of Geophysics and Planetary Physics and Department of Earth and Space Science
3845 Slichter Hall, Box 951567, 595 East Circle Drive
University of California, Los Angeles, California 90095

**Abstract**: We report new tests of the critical earthquake concepts performed on rockbursts in deep South African mines. We extend the concept of an optimal time and space correlation region and test it on the eight main shocks of our catalog provided by ISSI. In a first test, we use the simplest signature of criticality in terms of a power law time-to-failure formula. Notwithstanding the fact that the search for the optimal correlation size is performed with this simple power law, we find evidence both for accelerated seismicity and for the presence of logperiodic behavior with a prefered scaling factor close to 2. We then propose a new algorithm based on a space and time smoothing procedure, which is also intended to account for the finite range and time mechanical interactions between events. This new algorithm provides a much more robust and efficient construction of the optimal correlation region, which allows us the use of the logperiodic formula directly in the search process. In this preliminary work, we have only tested the new algorithm on the largest event on the catalog. The result is of remarkable good quality with a dramatic improvement in accuracy and robustness. This confirms the potential importance of logperiodic signals. Our study opens the road for an efficient implemention of a systematic testing procedure of real-time predictions.

**1-Our tribute to Leon**

It is a great pleasure and honor to participate to Leon's 75th birthday celebration and show our immense appreciation for his depth and breadth. The second author, a statistical physicist coming to the field of seismology and earthquake modeling in 1987, has benefited considerably from Leon's insights and rigor in addressing the difficult problem of earthquake modeling. Since the day we started our scientific relationship during a memorable conference in Cargese, Corsica in 1988, we have spent considerable time together, discussing and arguing about the key physical ingredients for understanding the complex organization of earthquakes. This was the time when Leon became very interested with the potential application of the emerging concepts of self-organization. During our learning period in which Leon has played an instrumental role, the more we read on seismology and earthquake modeling, the more we found that Leon has contributed significantly in essentially all the important problems, as reflected in the eclectic contributions to this special volume. We have followed his footprints with pleasure and inspiration. The second author feels particularly privileged to have shared these many hours of intense discussions. They have shaped irreversibly his approach to this field and also to science in general.

**2-The critical earthquake concept**

The first concrete fruit of our collaboration with Leon was a statistical model of earthquake foreshocks (*Sornette et al.,* 1992), which extended to the domain of strongly heterogeneous systems previous modeling effort by Yamashita and Knopoff (1989). This model is based on a realistic model of dynamically evolving damage (*Sornette and Vanneste*, 1992) which produces many growing interacting micro-cracks with an organization which is a function of the damage-stress law. Under a step-function stress loading, the total rate of damage, as measured for instance by the elastic energy released per unit time, increases on average as a power law of the time-to-failure. In this model, rupture is a genuine critical point and occurs as the culmination of the progressive nucleation, growth and fusion between microcracks, leading to a fractal network of cracks. This simple model has since then been found to describe quantitatively the experiments on the electric breakdown of insulator-conducting composites (*Lamaignère et al.,* 1996) and the damage by electromigration of polycrystalline metal films (*Bradley and Wu*, 1994). This led use to propose and test on real engineering composite structures the concept that failure in fiber composites is a genuine a "critical" point (*Anifrani et al.,* 1995). This critical behavior may correspond to an acceleration of the rate of energy release or to a deceleration, depending on the nature and range of the stress transfer mechanism and on the loading procedure. We were thus led to propose that the power law behavior of the time-to-failure analysis should be corrected for the presence of log-periodic modulations (*Anifrani et al.,* 1995). Log-periodicity is the signature of a hierarchy of characteristic scales in the rupture process. Since then, this method has been used extensively during our continuing collaboration with the French Aerospace company Aérospatiale on pressure tanks made of kevlar-matrix and carbon-matrix composites embarked on the European Ariane 4 and 5 rockets. In this application, the method consists in recording acoustic emissions under constant stress rate and the acoustic emission energy as a function of stress is fitted by the above log-periodic critical theory. One of the parameter is the time of failure and the fit thus provides a ``prediction'' when the sample is not brought to failure in the first test (*Anifrani et al.,* 1995). The results indicate that a precision of a few percent in the determination of the stress at rupture is typically obtained using acoustic emission recorded about 20% below the stress at rupture. We now have a better understanding of the conditions, the mathematical properties and physical mechanisms at the basis of log-periodic structures (*Sornette*, 1998).

At the same time, it became enticing (*Sornette and Sammis*, 1995) to apply similar considerations to earthquakes. Indeed, over the years there has been a growing evidence that a significant proportion of large and great earthquakes are preceded by a period of accelerating seismic activity of moderate-sized earthquakes. These moderate earthquakes occur during the years to decades prior to the occurrence of the large or great events and over a region much larger than its rupture zone. Theoretically, the critical earthquake concepts was first formulated by Vere-Jones (1977) using critical branching models and Allègre et al. (1982) using the percolation model and real-space renormalization group (see Sornette (2000) for a discussion). The russian school has also extensively developed this concept  (*Keilis-Borok*, 1990). Sornette and Sammis, (1995) identified a specific measurable signature of this criticality (see also (*Sornette and Sornette*, 1990)) in terms of a power law acceleration of the Benioff strain previously interpreted as an exponential mechanical damage rate (*Sykes and Jaumé*, 1990; *Bufe and Varnes*, 1993). The combined observational and simulation evidence now seems to confirm that the period of increased moment release in moderate earthquakes signals the establishment of long wavelength correlations in the regional stress field. Large or great earthquakes appear to dissipate a sufficient proportion of the accumulated regional strain to destroy these long wavelength stress correlations. They can thus be considered as different from smaller earthquakes. According to this model, large earthquakes are not just scaled-up version of small earthquakes but play a special role as "critical points*" (Bowman et al.*, 1998; *Jaume and Sykes*, 1999). These later works were inspired by the empirical observation of Knopoff et al. (1996).

This critical earthquake concept has been further strengthened by showin that the nature and strength of heterogeneity controls the rupture process: increasing the disorder changes rupture from first-order (abrupt) to critical (continuous with power law properties) (*Andersen et al.,* 1997). This was anticipated early by Mogi (1969), who showed experimentally on a variety of materials that, the larger the disorder, the stronger and more useful are the precursors to rupture. For a long time, the Japanese research effort for earthquake prediction and risk assessment was based on this very idea (*Mogi*, 1995).  In our two-dimensional spring-block model of surface, inspired by the famous Burridge-Knopoff model (*Burridge and Knopoff*, 1967), the stress can be released by spring breaks and block slips (*Andersen et al.,* 1997). This spring-block model may represent schematically the experimental situation where a balloon covered with paint or dry resin is progressively inflated. An industrial application may be for instance a metallic tank with carbon or kevlar fibers impregnated in a resin matrix wrapped up around it which is slowly pressurized (*Anifrani et al.,* 1995).  As a consequence, it elastically deforms, transferring tensile stress to the over-layer. Slipping (called fiber-matrix delamination) and cracking can thus occur in the over-layer. In the presence of long-range elasticity, disorder is found to be always relevant leading to a critical rupture. However, the disorder controls the width of the critical region (*Sornette and Andersen*, 1998). The smaller it is, the smaller will be the critical region, which may become too small to play any role in practice. Numerical simulations of Sahimi and Arbati (1996) have confirmed that, near the global failure point, the cumulative elastic energy released during fracturing of heterogeneous solids with long-range elastic interactions exhibit a critical behavior with observable log-periodic corrections. The presence of log-periodic correction to scaling in the elastic energy released has also been demonstrated numerically for the thermal fuse model (*Johansen and Sornette*, 1998) using a novel averaging procedure, called the "canonical ensemble averaging". Recent experiments on the rupture of fiber-glass composites has also confirmed the critical scenario (*Garcimartin et al.,* 1997).

This paper presents a novel study testing this critical point concept for rockbursts occurring in deep South African mines. This system offers an intermediate range of scales between the laboratory and the earth crust. The quality of the data is very high, for instance the hypocentral depth are determined very accurately to within a precision of a few meters as seismographs are placed at depth in the mining tunnels. The data has been kindly provided by ISSI, Welkom, South Africa and we thank A. Mendecki and W. de Beer for this and for discussions.

### 3-Description of the data set

The catalog provided by ISSI lists the source parameters of 2487 events in a domain of half-size 270m in its horizontal and transverse dimension and unlimited in the vertical direction. These events have been recorded over a period from 97/02/01 to 97/09/30 by five stations or more in a deep South African mine in the Welkom region during active mining. The mining activity is the driving force which modifies the stress field and triggers small earthquakes, the so-called rockbursts. The source parameters are respectively the occurrence time, the 3D spatial location of the hypocenter, the seismic moment, and the energy radiated through seismic waves. Eight events were selected to test the concept of critical acceleration of seismic energy release before their occurrence. Those events were simply chosen as the 8 largest shocks, with moment larger or equal to $10^{12}$ Nm. Figures 1a and 1b show the spatial location of these events, together with the rest of the catalog, in horizontal and vertical projections. The so-called " large events " are figured with diamond symbols to distinguish them from other shocks. The whole seismicity pattern reveals the existence of roughly two main clusters striking about N50°, and lying mostly between 1800 and 2000m in depth. The parameters of the selected large events are listed in Table 1. The first column gives the label of the event; the second one gives the date and time of occurrence; the third column gives the normalized time used for computations: 1997, January 1st, 0h00 corresponds to time=0, whereas 1997, December 31, 24h00 corresponds to time=1. The very first event in the original catalog occurred on February $1^{st}$ 1997 (time=0.085), whereas the very last one occurred on September 29 1997 (time=0.745); the 3 following columns give the spatial location; the next one gives the released seismic moment; the last one shows the ranking of the 8 events according to their seismic moment. A simple visual inspection of Figure 1 shows that 4 of the selected events (labelled respectively as 2, 3, 4 and 5) are not located within any of the two main clusters. Their shortest distance to such a cluster ranges from 100 to 300m.

**Table 1**

| Event # | Date and time | Normalized time | X (m) | Y (m) | Z (m) | Seismic moment ($10^{12}$ Nm) | Rank |
|---------|---------------|-----------------|-------|-------|-------|------------------------------|------|
| 1 | February 5 16h30'12'' | 0.0977 | 6263 | 7951 | -1876 | 2.8 | 3 |
| 2 | April 3 19h35'54'' | 0.254 | 6719 | 7905 | -1954 | 1.4 | 5 |
| 3 | April 6 01h13'12'' | 0.260 | 6312 | 8121 | -1795 | 13.5 | 1 |
| 4 | May 4 18h35'31'' | 0.339 | 6329 | 8028 | -2278 | 9.4 | 2 |
| 5 | May 26 23h47'26'' | 0.399 | 6603 | 7849 | -2194 | 1.0 | 8 |
| 6 | June 6 14h02'45'' | 0.429 | 6475 | 7823 | -2019 | 2.8 | 4 |
| 7 | June 23 17h11'27'' | 0.476 | 6284 | 8016 | -1931 | 1.2 | 7 |
| 8 | August 15 22h45'14'' | 0.622 | 6591 | 7825 | -1860 | 1.3 | 6 |

**4- Theoretical formulae and procedure**

*4.1 Quantification of seismic energy release*

The aim of this work is to check for the existence of a critical behaviour in the seismic strain energy release pattern before a large event. This critical behaviour has to be found in the spatial and temporal pattern of the rest of the catalog, the so-called "background seismicity". Several parameters can be computed from events source parameters, among which we can list the seismic moment, the radiated seismic energy, the apparent volume and stress, the rupture surface or length, etc. In this preliminary study, we shall use the methodology that revealed such critical patterns in the precursory activity of large crustal scale shocks (*Bowman et al.,* 1998). This method focuses on some power $q$ of the scalar seismic moment $M_0$. Let be $t_0$ the origin time of the catalog used to study the precursory activity before a large event occurring at time $t_c$. This means that the only events that are taken into account occur within $[t_0 ; t_c[$. We consider that, excluding the large event itself, we deal with a set of $N$ such events of seismic moment $M_{0k}$, $k=1,...,N$, occurring at discrete times $t_k$. We define a generalized cumulative strain energy release function, defined as:

$$S(t) = \sum_{k=1}^{N} \int_{0}^{t} M_{0k}^{q} \delta(u - t_k) du \qquad \text{(eq. 1)}$$

where $\delta(t)$ is the Dirac function and $t$ is lower or equal to $t_c$. If $q=0$, each event has the same weight in eq. 1, so that $S(t)$ is simply the cumulated number of events as a function of time $t$. In the other hand, if $q=1$, $S(t)$ is the cumulated released seismic moment and is mainly sensitive to the largest events, which are also the rarest ones. This results in a poor statistical significance. In practice, one has to choose an exponent between 0 and 1. Taking $q=1/3$, and assuming strict scale invariance of the stress drop with event size, gives a quantity proportional to the length of the fault which ruptured or, equivalently, to the coseismic displacement along this fault. Taking $q=2/3$ gives a quantity proportional to the rupture plane area. We shall choose here an intermediate quantity $q=1/2$. This gives the so-called *Benioff strain* (neglecting some prefactors to integral (1) which depend on stress drop and elastic properties of the medium). Our results reported below remain robust if we use q=1/3. They change however and become unreliable if we use q=2/3 which is already too large and give too much weights to the largest events.

### 4.2 Critical acceleration before the main shock

The critical behavior of background seismicity can be expressed as a power-law acceleration of the cumulative Benioff strain release before a large event, characterized by a critical exponent (which can be real or complex), the singularity being localized at time=$t_c$, the occurrence time of the large shock. Let us first consider the case when the exponent is real. The critical Benioff strain release takes the form:

$$\widetilde{S}(t) = A + B(t_c - t)^z \qquad \text{(eq. 2)}$$

with $t<t_c$, and $z$ is chosen within *[0;1]*. This choice insures that the total released strain (including the main shock) remains finite, but that its derivative according to time diverges at $t=t_c$. The 3 unknown parameters are $A$, $B$ and $z$, but there are 2 other hidden unknown variables, which control the data set used to compute $S(t)$. Indeed, it is legitimate to think that all the events listed in the catalog are not precursors of all the large shocks. For instance, is the very first event of the catalog (with seismic moment 7.4 $10^8$ Nm) a precursor of the large event #8? For a given large event, the first hidden unknown parameter is thus $t_0$ which marks the temporal beginning of the " useful " part of the catalog, *i.e.* the time interval within which precursors occur. The second hidden unknown is a spatial one. The large event is the culmination of a collective process which occurs within a critical domain. It is reasonable to think that this zone has a finite size (in fact, it can not be larger than the total extent of the catalog). In a first approximation, we shall assume that the large event is located at the center of this critical domain, which is considered as spherical with radius $R_c$, here following Bowman et al. (1998). This approximation has the advantage of being very parsimonious. This will be relaxed if future investigations as there is absolutely no reason for the critical domain to be either symmetric, nor centered on the main shock.

To test the critical point hypothesis for a given large event, we shall thus have to determine the size of the critical zone both in space and time (*i.e.* $R_c$ and $t_0$), then to compute $S(t)$ using eq. 1, then to fit this function using eq. 2 to find the explicit unknowns $A$, $B$ and $z$. The critical point will then be fully characterized. But we have still to show that eq. 2 provides a statistically significant description of a given data set, in contrast to other, possibly non-critical behaviors. We thus first use eq. 2 to fit $S(t)$ and compute the empirical variance:

$$V_p = \sum_{k=1}^{N} \left( \tilde{S}(t_k) - S(t_k) \right)^2 \qquad \text{(eq. 3)}$$

The same data set (*i.e.* corresponding to the same $R_c$ and $t_0$) is then fitted using a simple linear relationship. A new empirical variance $V_l$ is computed, as well as the ratio $r=V_p/V_l$. If $r$ is close to 1, a linear fit explains the data behavior as well as a power law, making it difficult to argue for the critical point hypothesis. In the other hand, if $r$ is significantly lower than 1, then the use of eq. 2 is a significant improvement in describing the data compared to the use of a simple linear relationship. This implies a large confidence in the critical point hypothesis. However, the results must be discussed in the light of another parameter, which is the size of the data set used to perform the analysis. In order to get a reliable statistical significance, we shall not consider data sets containing less than 7 events.

To determine the 5 unknowns of our problem, we use the following procedure:
(1) fix $R=0$.
(2) fix $t_0=0$.
(3) select in the whole catalog all events occurring within a distance $R$ of the main event, and within $[t_0;t_c[$. If the number of events is too low (for example lower than 7 in this present work), go to step 10
(4) compute the Benioff strain function using eq. 1.
(5) fit the Benioff strain function using eq. 2 and compute $V_p$.
(6) fit the same function using a linear function, and compute $V_l$.
(7) compute the ratio $r=V_p/V_l$.
(8) increment $t_0$ by $\Delta t_0$ and go to step (3) until $t_0=t_c$
(9) store the $t_0$ value for which $r$ is minimal, as well as the corresponding $r$ and $z$ values.
(10) increment $R$ by $\Delta R$ and go to step (2) until $R$ reaches the spatial size of the catalog (about 1000m)
(11) end

Here, we take $\Delta R=1$m, and $\Delta t_0=0.01$ (about 3.25 days). These choices do not result from any theoretical justification, but ensure a good compromise between accuracy and computation time. This procedure is repeated for each main event (*i.e.* each value of $t_c$). We can then draw several curves concerning each large shock. The first one represents $r$ as a function of $R$. Parts of the curve where $r$ is minimal help to define the value of $R_c$ (see example on **Figure** 2a). If several minima are obtained, we choose the one for which the variation of $r$ with $R$ is the slowest, defining the more stable minimum. Another curve shows the best value of $z$ as a function of $R$. The value of $z$ at $R=R_c$ defines the so-called critical exponent (see Figure 2b for an example). The optimal cumulated Benioff strain function obtained for $R=R_c$ can be drawn as a function of time (see Figure 3 for example), showing an acceleration before $t_c$. Data often display some oscillations around the theoretical power-law, which can be reminiscent of the log-periodic signature of an underlying discrete scale invariance, *i.e.*, the exponent in eq. 2 is complex-valued. This leads to the introduction of some log-periodic corrections to a simple power-law. The optimal data set corresponding to $R=R_c$ is then fitted with the following formula:

$$\bar{\bar{S}}(t) = A + B\left(t_c - t\right)^z \left( 1 + C\cos\left( 2\pi \frac{\log\left(t_c - t\right)}{\log \lambda} + \phi \right) \right) \qquad \text{(eq. 4)}$$

where the unknowns are $A, B, z$ (a real critical exponent), $C, \lambda$ (the natural contraction factor of discrete scale invariance), and $\phi$ (a phase angle). It is important here to mention another simplification: we do not repeat any optimization over $R_c$ and $t_0$ to find the unknowns of eq. (4): the data set used for this last fit is the data set that was previously optimized for a simple power law fit. The reason of this choice is that a new optimization for the more complicated eq (4) would require too much computation time that we had at our disposal in this preliminary study. Section 7 presents a new methodology which reduces significantly the computation time and gives very encouraging results. We must acknowledge here that the optimal data set for eq (2) is not necessarily the optimal data set of eq (4), but we shall suppose it approximately true. Logperiodic fit was not performed with data sets containing less than 14 events.

**5- Application to rockburst data**

We now present the results obtained for the precursory behavior of background seismicity before each of the events shown in Table 1.

*1 - Event # 1*

The $(R;r)$ curve is presented on Figure 2a. Two minima can be defined. The first one is located at $R=144$m (with $r=0.41$), the second one at $R=297$m ($r=0.22$). This last one is located within a wider "basin" than the previous one, and is characterized by a smaller $r$ ratio. This is why we choose $R_c=297$m. figure 2b shows the variation of the determined $z$ exponent with $R$. Within the previously defined basin minimum, the $z$ exponent is of the order of 0.3 to 0.4. Its value at $R_c$ is 0.27. The value of $t_0$ defining the optimal data set corresponds to the beginning of the catalog (note here that $t_c$ is close to the occurrence time of the very first event). Thus, the time window $t_c$-$t_0$ is 4.6 days, which is the maximum size allowed by the catalog for this event, and thus constitutes a lower bound. Figure 6 finally presents the optimal $S(t)$ function (called

Generalized Energy), as well as the associated fitting curve. The last point on the right corresponds to the main event, but we have to remember that this one was not used during the fitting procedure. It is plotted here only to compare with the last point of the fitting curve, which gives an idea of the error done on the prediction of the size of the main event (whereas its occurrence time $t_c$ is supposed to be known *a priori*). In this case, the predicted Benioff strain of the main event is lower than its actual size. Such a discrepancy can be expected if we consider that the main event will largely depend on finite-size effects which have non-universal properties. Thus, the prediction of the size of the large rockbursts is not expected to be reliable by using only the time-to-failure analysis. To improve, we would need to develop a more precise spatial analysis incorporating the tensorial aspect of the rupture. Figure 4 presents the same data as in Figure 3, fitted using eq. 4. The variance we obtain is 3 times lower than with a simple power law, which is a significant improvement. The new $z$ exponent is found equal to 0.4, whereas we have $C$=0.05 and $\lambda$=2.2. This value for $\lambda$ is in good agreement to the value expected from previous works and theoretical arguments (*Sornette*, 1998).

### *2 - Event # 2*

This event is located outside the two main spatial clusters. Figure 5 shows the $(R,r)$ curve which defines a basin minimum around $R_c$=303m. The $r$ value is then equal to 0.18. The $t_0$ value is 0.25, such that the total time window available is the smallest that can be defined for this event (about 1.6 days). Figure 6 shows the $z$ value as a function of $R$. Its value at $R_c$ is the smallest that can be obtained by our inversion, *i.e.* 0.01. This small value of the exponent means intuitively that $S(t)$ is approximately linear and then abruptly increases close to the main shock. Figure 7 shows the data set with the fitting curve. We performed a second analysis, forcing the $z$ exponent to be larger than 0.2. The new value we determined was 0.2, which is again the lowest bound allowed. According to these results, we must acknowledge that no truly convincing critical acceleration could be detected before this event using a simple power-law.

We performed a fit of the data shown on Figure 7 with the log-periodic formula. The $z$ exponent was again found equal to 0.01, whereas $C$=0.01, and $\lambda$=1.4, the variance being 73% of the variance obtained with a simple power law. The fit is presented on Figure 8. We performed a second fit by forcing $\lambda$ to be within *[1.5;3]*. We found the same $z$, $C$ and variance values, whereas $\lambda$ was found equal to 2.8. Thus, the data can not be reasonably explained by a logperiodic behavior, when using the same correlation size R derived from the simple power law. It is possible, in view of the logperiodic structures observed in figure 8, that a direct search with formula (4) would be successful. We will address this improved methodology in the last section.

### *3 - Event # 3*

This event is also located outside of the two main clusters. The $(R;r)$ curve (Figure 9) is rather complex. We study in more details the 3 best minima, located respectively at $R_c$=225m, 407m and 487 m. Their respective ratios $r$ are 0.276, 0.195, 0.174. One can observe a relatively large basin minima within the interval [400m;500m]. Figure 10 shows the ($R$;$z$) curve, which is even more complex. The values of $z$ relative to the 3 minima are respectively 0.65, 0.57, 0.48. The time windows are respectively 37 days, 11 days and 11 days. Figure 11 shows the data and theoretical fit for $R_c$=487m. We fitted those 3 data sets with the logperiodic formula. The results are:

$R_c$=225m: $z$=0.71, $C$=0.04, $\lambda$=2.1. The variance is 63% of the one obtained using a simple power law.

$R_c$=407m: $z$=0.60, $C$=0.04, $\lambda$=2.5. The variance is 69% of the one obtained using a simple power law.

$R_c$=487m: $z$=0.48, C=0.04, $\lambda$=2.2. The variance is 67% of the one obtained using a simple power law (see fit on Figure 12).

These results show that a simple powerlaw acceleration is significant and the logperiodic correction brings an improvement to the data description which is not as significant as it is the previous cases. The simplest working description is the simple power law with Rc=450m, z=0.5, and a time windows of 11 days. We note however that the value of the prefered scaling ratio $\lambda$ value close to 2 is significant and in full agreement with event #1 and also with our many experiences on other systems [*Sornette*, 1998].

### 4 - Event # 4

This event is also located outside the main clusters of activity. The (R;r) curve (Figure 13) defines a rather small basin around R=400m. The minimum occurs for Rc=404m. The corresponding z value (see Figure 14) is 0.51, and r=0.30. Figure 15 shows the associated data set, defined within a time window of 51 days. The logperiodic formula yields z=0.55, C=0.01, $\lambda$=1.7 (see Figure 16), whereas the variance is 72% of the variance obtained with a simple power law. The value of $\lambda$ is again close to 2. However, the statistical significance of the logperiodic formula to describe this data set is not very large as the improvement on the residue is not as large as in event #1. But again, let us keep in mind that the optimization of the correlation radius R has been performed with the simple power law and a direct search with the logperiodic formula may improve significantly the results.

### 5 - Event # 5

This event is the last one which is located outside the main clusters of activity. Figure 17 shows that a single minimum can be defined in the (R;r) curve. It occurs at Rc=247m, with r=0.17 (and a time window of 18 days). However the basin is very narrow, which may signal an instability. Figure 18 shows the corresponding z values. In this case, only 7 points define the optimal data set. For all other R values, r is close to or larger than 0.5, which means that a power law does not describe much more significantly the data than a simple linear behavior. We didn't perform any logperiodic fit with this optimal data set because of the limited number of events. However, we had a closer look to the second best minimum which occurs at Rc=377m (time window=14.6 days). Performing a logperiodic fit on this new dataset did not give any satisfying result. However, moving tc from its real value of about 0.4 (see Table 1) to 0.395 (that is by advancing its occurrence time by 1.8 days) gave the following results for the logperiodic fit (see Figure 19): z=0.275, C=0.03, $\lambda$=2.4. A logperiodic behavior is thus acceptable if we assume a little uncertainty on tc (in this case, the main event appears as slightly advanced in time). The variance of the logperiodic fit is only 11% of the one of a linear fit and thus provides a very significant improvement. It is thus very significant.

Let us comment briefly on this procedure of adjusting tc. First, in a predicting mode, tc would not be fixed to the time of the main event as done presently but would be an adjustable variable that the fitting procedure must determine. Second, the underlying model in terms of a critical point predicts that the system-specific time-of-rupture is in general slightly moved away from what it would be in an asymptotically large system. This phenomenon is known under the concept of finite-size-scaling and is particularly important to account for in the presence of logperiodic oscillations (*Johansen and Sornette*, 1998). Thus, there is a good theoretical justification for the expectation that the power law and the logperiodic power law fits will critical times tc that are slightly misplaced. This error is intrinsic to the problem and constitutes a fundamental limitation.

### 6 - Event # 6

Two minima are chosen: the first one occurs at R=86m, the second one at R=223m (see Figures 20 and 21). The associated data sets contain respectively 13 and 61 events, whereas the time windows are respectively 25 days and 7 days. We choose Rc=223m because the z value is in better agreement with other results. A comment is in order here: for our accumulated experience, we reject fits that give an exponent that are either too small (smaller than 0.1) or too large (larger than 0.9). For Rc=223m, we obtain r=0.203, z=0.54 (see Figure 22) (for R=86m we get r=0.129 and z=0.13). The logperiodic fit is performed for $R_c$=223m. We obtain z=0.55, C=0.05 and $\lambda$=1.5. The variance is 71% of the variance obtained from a simple power law fit. However, using the same data set without the points for which t<0.412, we obtain: z=0.48, C=0.05 and $\lambda$=2.1 and a much better fit (see Figure 23). This simple case shows that minor changes in the dataset can lead to a better determination of a logperiodic behavior (in this case, the variance of the logperiodic fit is less than 9% of the variance of the linear fit, which thus leads to a very significant result). This « data picking » is in fact warranted by the fact that the best time window has been determined by using the simple power law. As already mentioned, a direct optimization with the logperiodic formula (4) would probably have modified it significantly. Our rejection of data points prior to *t* =0.412 for the logperiodic fit is thus a simple way to adjust the procedure. Note again the excellent value found for $\lambda$ which is very close to 2, reinforcing the statistical significance for the existence of a discrete scale invariance in the mine rockbursts.

### *7 - Event # 7*

The ($R;r$) curve shows a very wide basin minimum for $R>400$m (see Figure 24). However, the corresponding $z$ exponent is equal to 0.01 (Figure 25) which is not reasonable. Moreover the corresponding time window is the shortest one that is allowed in this computation, which is about 2 days. The two minima we finally retain are located at $R=40$m and $R=80$m, which are rather short distances. For $R=40$m, we obtain $r=0.09$, $z=0.63$, but only 7 points define the data set in a time window of 57 days. For $R=80$m, we obtain $r=0.26$, $z=0.53$ and we have 52 data points for a 64 days time window (Figure 26). The logperiodic fit is performed only for Rc=80m, and we get z=0.64, C=0.08, $\lambda$=5.1, for a variance which is 51% of the one obtained with a simple power law (Figure 27).

### 8 - Event # 8

A single reasonnable minimum is obtained at $R_c$=31m (Figure 28). The value of $r$ is equal to 0.25. For every other minimum, $r$ is larger than 0.5. The value of $z$ we determined is 0.09 (Figure 29). The time window is the maximum value allowed, which is 195 days. The fit is presented on Figure 30. The logperiodic formula gives the following parameters: $z$=0.43, $C$=0.08, $\lambda$=9.5, and the variance is 51% of the variance we obtained with a simple power law. If we consider for this same data set only the events occurring after $t$=0.25, the logperiodic fit gives: $z$=0.43, $C$=0.09, $\lambda$=2.5 (see Figure 31), with a $\lambda$ value much more in agreement with the previous results (the variance of the logperiodic fit being 22% of the one using a simple linear law which is very significant). The adjustement of the time window for the logperiodic fit performed on the data set optimized to the simple power law is done again as discussed above.

## 6- Discussion

Table 2 presents a recapitulation of the results. The different columns contain respectively the label of the main event (from 1 to 8), the $z$ and $\lambda$ values obtained from the fits, the size $R_c$ of the critical domain, the duration of the time window in which Benioff strain acceleration before the event is detected, the ranking of the event according to its size (from 1 to 8), the location in or out of one of the two main clusters of activity (" in " or " out ") described in the introduction and the occurrence of a significant acceleration of the Benioff strain before the main event (" yes " or " no "). A $\lambda$ value followed by the symbol "?" means that the logperiodic fit was not found to improve significantly a simple power law fit.

**Table 2**

| Event # | $z$ | $\lambda$ | $R_c$ | Time window (days) | size rank | within a cluster ? | acelerating behavior ? |
|---------|-----|-----------|-------|--------------------|-----------|--------------------|------------------------|
| 1 | 0.4 | 2.2 | 297m | > 4.6 | 3 | in | yes |
| 2 | 0.01 | 1.4 | 303m | <1.6 | 5 | out | no |
| 3 | 0.50 | 2.3 ? | 450m | 11 | 1 | out | yes |
| 4 | 0.51 | 1.7 ? | 404m | 51 | 2 | out | yes |

| 5 | 0.28 | 2.4 | 377m | 14.6 | 8 | out | yes |
|---|------|-----|------|------|---|-----|-----|
| 6 | 0.48 | 2.1 | 223m | 7 | 4 | in | yes |
| 7 | 0.64 | 5.1 | 80m | 64 | 7 | in | yes |
| 8 | 0.43 | 2.5 | 31m | >195 | 6 | in | yes |

This table shows several interesting features:

- all events except #2 show a significant precursory acceleration of Benioff strain release. Event 2 is the 5[th] event in size and is located on the edge of the hypocenters pattern. As our methodology implies the existence of an approximative symmetry (through the introduction of a spherical critical domain), the event does not fulfill the most favorable conditions for our study.

- Among the events preceded by an acceleration, only events #3 and 4 show that a logperiodic behavior does not describe the data more efficiently than a simple power law. Despite the fact that these events are the two biggest ones in the catalog, one should note that both are, as event #2, located outside of the two main clusters of activity. A similar conclusion also holds for event #5, which is the smallest of the main events, but which displays a critical behavior. However, we must remember that we had to slightly modify the time interval of the fit of event 5 to obtain a significant logperiodic behavior. This highlights the fact that is would be useful here to perform an optimization on $R_c$ and $t_0$ using the full logperiodic formula and allow a little uncertainty on $t_c$. We can also imagine to consider arbitrarily shaped critical domains.

- the exponent $z$ is most often found within [0.3;0.6], in good agreement with the values found previously with other datasets (*Anifrani et al.*, 1995; *Bowman et al.,* 1998).

- The contracting factor $\lambda$ is most often close to 2, which is again in good agreement with other results on discrete scale invariance (*Sornette*, 1998). Only event #7 is an exception, but we should consider the fact that it is only the 7[th] larger shock in the catalog, so that its signature may not be very strong. As it was possible to find a critical signature for event 5, which is smaller, it would be interesting to perform a complete optimization using the full logperiodic formula  for the event (see above).

- The radius of the critical domain is of the order of 300-400m for most events. This implies that the total width of the critical domain is 600-800m, which is comparable to the spatial size of the catalog. Only events #7 and 8 have much smaller radii, which is perhaps due to their smaller size. Larger events should be studied in the future to check the relationship between the size of the main event and the size of the critical domain, as proposed in (*Bowman et al.*, 1998; *Jaumé and Sykes*, 1999).

- The time window, which defines the temporal size of the critical domain, is of the order of a few tens of days. Only events # 2 (not critical) and 8 have very different time characteristics. For this last event, it is abnormally large (more than 6 months).

## 7- An optimization algorithm for the logperiodic formula

The algorithm proposed in section 4 suffers from a major shortcoming, which is that is is not able to capture the mechanics of successive ruptures. For exemple, it is clear from mechanical consideration that, for a given foreshock, the closer it occurs to the future mainshock, the largest is its influence on that main event. This results from the effect of stress redistribution after an earthquake (*King et al.,* 1994). This leads us to define weights for each precursory event, depending on its space and time distance to the future large event. In addition, a very important advantage of such a weighting is that it will suppress the edge effect introduced by our previous definitions of $R$ and $t_0$ : the algorithm of section 4 implies that the weight of an event in the Benioff strain function is proportional to the square root of its seismic moment, except if it is located outside of the zone defined by ($R;t_0$), in which case its weight is 0. It is clear that this leads to an instability: increasing $R$ by a small amount may suddenly include new events that may then modify significantly the results. This sensitivity is not justifiable physically.

We shall here introduce a smoother variation of weight with distance. We define the cumulative Benioff strain (parameterized by $R$ and $t_0$) as:

$$S(t, t_0, R) = \sum_{k=1}^{N} \int_0^t M_{0k}^{1/2} F(k, R, t_0) \delta(u - t_k) du \qquad \text{(eq 5)}$$

with

$$F(k, R, t_0) = \exp\left(-\frac{1}{2}\left(\frac{D_k}{R}\right)^2\right) \exp\left(-\frac{t_c - t_k}{t_c - t_0}\right) \qquad \text{(eq 6)}$$

$N$ is now the total number of events within the catalog, $D_k$ is the spatial distance between the $k^{th}$ event and the main shock. The function $F(k,R,t_0)$ is a smoth function of both spatial and temporal coordinates. The Gaussian in eq.(6) replaces the indicator function used previously, which was equal to one within the disk and zero outside. Note that this previous case is exactly retrieved in the limit where we replace the exponent 2 by a very large value. This demonstrates that the Gaussian is only introduced as a convenient smoothing device that should ensure both a better stability and the possibility to use a coarser sampling, hence leading to a faster algorithm. The exponential time function in eq. (6) is also introduced to provide a smoothing filter. It is also intended to capture the physical phenomenon of a possible ductile relaxation between events. The time $t_0$ will be optimized and corresponds to an optimization of the relaxation time $t_c - t_0$.

We can now define our new procedure, similar to the one defined in section 4, except that we now use equation 4 (logperiodic formula). The more robust cumulative Benioff strain function obtained by this filtering allows us to use a coarser sampling of $t_0$ and $R$ and thus authorizes fitting with the more complex logperiodic formula (4). Recall that in this first testing procedure, we suppose that $t_c$ is known and given by the time of the main shock. The algorithm is as follows:
(1) fix $R=10$m.

(2) fix $t_0$ close to 0.

(3) compute the Benioff strain function using eq. 5.

(4) fit the Benioff strain function using eq. 4 and compute $V_{lp}$.

(5) fit the same function using a linear function, and compute $V_l$.

(6) compute the ratio $r=V_{lp}/V_l$.

(7) increment $t_0$ by $\Delta t_0$ and go to step (3) until $t_0=t_c$

(8) store the $t_0$ value for which $r$ is minimal, as well as the corresponding $r$ and $z$ values.

(9) increment $R$ by $\Delta R$ and go to step (2) until $R$ reaches the spatial size of the catalog (about 1000m)

(10) end

Here, we take $\Delta R$=20m, and $\Delta t_0$=0.04 (about 2 weeks). We focus on event #3, which is the largest one. Figure 32 shows the optimal data set together with the logperiodic fit (which is barely visible on the figure, so good is the fit that the logperiodic curve goes through essentially all the data points). The optimal values found by this procedure are $R_c$=310m, $t_0$=1 month, $z$=0.49, $C$=0.02, and $\lambda$=2.0. The variance is about 2% of the one obtained with a simple linear fit i.e. the residue has been decreased by a factor of 50! This is a very significant result, strongly qualifying the logperiodic formula. We note also that the prefered scaling ratio $\lambda$=2.0 is very good. This test on the largest event of the ISSI catalog is thus very encouraging as this new procedure seems to greatly enhance our ability to describe the data.

**8-Conclusion**

We have presented first the concept of an optimal spatial and temporal correlation domain and have tested it on the eight main shocks of the catalog provided by ISSI of rockburst in deep South African miines. In a first test, we have used the simple power law time-to-failure formula. Notwithstanding the fact that the search for the optimal correlation size has been performed with this simple power law, we have found good evidence for the presence of logperiodic behavior with a scaling factor $\lambda$ close to 2. This led us to propose a new algorithm based on a space and time smoothing procedure, which is also intended to account for the finite range and time mechanical interactions between events. This new algorithm provides a much more robust and efficient construction of the optimal correlation region, which permitted the use of the logperiodic formula directly in the search process. Testing this procedure on the largest event on the ISSI catalog, we obtain a very good result. This seems to confirm the importance of the logperiodic signals that appear very strong, even if such signals are also known to occur by the very nature of fluctuations in power law signals (*Huang et al.,* submitted).For the future, we need to test this new procedure on other events. Secondly, we need to implement it for a real-time prediction. In this goal, we need to optimize the algorithm to make it faster for concrete use. Thirdly, the robustness and parsimony of the sampling open the possibility to use our improved logperiodic formulas (not described here), especially designed to account for the transition from non-critical to critical behavior.

**Figure 1a**
**Horizontal projection of the catalog**

**Figure 1b**
**Vertical projection of the catalog**

**Figure 2a**
**An example of the variation of the ratio $r$ with radius $R$ for event #1.**

**Figure 2b**
**Variation of the optimal $z$ exponent with $R$ for event #1.**

**Figure 3**
**Cumulative Benioff strain function and power law fit for the optimal data set of event #1 .**
**Note the existence of some oscillations around the power law fit.**

**Figure 4**
**Logperiodic fit of the optimal data set for event #1.**

**Figure 5**
**($R$;$r$) curve for event #2.**

**Figure 6**
**($R$;$z$) curve for event #2.**

**Figure 7**
**Power law fit of the optimal data set for event #2.**

**Figure 8**
**Logperiodic fit of the optimal data set for event #2.**

**Figure 9**
($R;r$) curve for event #3.

**Figure 10**
($R;z$) curve for event #3.

**Figure 11**
Power law fit of the optimal data set for event #3.

**Figure 12**
Logperiodic fit of the optimal data set for event #3.

**Figure 13**
($R;r$) curve for event #4.

**Figure 14**
($R;z$) curve for event #4.

**Figure 15**
Power law fit of the optimal data set for event #4.

**Figure 16**
Logperiodic fit of the optimal data set for event #4.

**Figure 17**
($R;r$) curve for event #5.

**Figure 18**
($R;z$) curve for event #5.

**Figure 19**
**Logperiodic fit of the optimal data set for event #5.**

**Figure 20**
**($R;r$) curve for event #6.**

**Figure 21**
**($R;z$) curve for event #6.**

**Figure 22**
**Power law fit of the optimal data set for event #6.**

**Figure 23**
**Logperiodic fit of the optimal data set for event #6.**

**Figure 24**
**($R;r$) curve for event #7.**

**Figure 25**
**($R;z$) curve for event #7.**

**Figure 26**
**Power law fit of the optimal data set for event #7.**

**Figure 27**
**Logperiodic fit of the optimal data set for event #7.**

**Figure 28**
**($R;r$) curve for event #8.**

**Figure 29**
(*R*;*z*) curve for event #8.

**Figure 30**
Power law fit of the optimal data set for event #8.

**Figure 31**
Logperiodic fit of the optimal data set for event #8.

**Figure 32**
Logperiodic fit obtained by the new optimization procedure
defined in section 6 for event #3.